\documentclass[12pt]{article}
\usepackage{latexsym}
\begin{document}

\vskip 3truecm

\begin{center}
{\large {\bf NEUTRONLESS $^{10}$Be-ACCOMPANIED TERNARY FISSION
 OF $^{252}$Cf}}
\end{center}

\vskip 2truecm

 A. S\u andulescu$^{1,2,3,4,5)}$, F. C\^ arstoiu$^{1,2,5)}$, 
 \c S. Mi\c sicu$^{1,2,5)}$, A. Florescu$^{1,3,4)}$,

 A.V. Ramayya$^{3,4)}$, J.H. Hamilton$^{3,4)}$, J.K.Hwang$^{3,4)}$, and 
 W. Greiner$^{3,4,5)}$

\vskip 2truecm

 $^{1)}$ Institute of Atomic Physics, Bucharest, P.O.Box MG-6, Romania 

 $^{2)}$ Gessellschaft f\"ur Schwerionenforschung mbH, 64291
         Darmstadt, 

~~~~Wixhausen, Germany
 
 $^{3)}$ Physics Department, Vanderbilt University, Nashville, TN 37235, USA

 $^{4)}$ Joint Institute for Heavy Ion Research, Oak Ridge, TN 37831, USA

 $^{5)}$ Institut f\"ur Theoretische Physik der J.W.Goethe Universit\"at,

 D-60054, Frankfurt am Main, Germany

\vskip 2truecm

\ { \bf Abstract } : A new type of decay corresponding to the
 neutronless $^{10}$Be-accompanied fragmentation
 of $^{252}$Cf is studied. We employ a cluster model similar to the 
 model used for the description of cluster radioactivity.
 No preformation factors were considered.
 The ternary
 relative isotopic yields were calculated as the ratio
 of the penetrability of a given ternary fragmentation over 
 the sum of penetrabilities of all possible ternary neutronless 
 fragmentations. The
 corresponding barriers between the light and heavy fragment
 and between the $^{10}$Be cluster and the two heavier fragments
 were computed with the help of a double folding potential generated
 by M3Y-$NN$ effective interaction and realistic fragment ground state
 deformations.
Also, we studied the influence of the fragment excitation energies
on the yields, by including the level densities and the 
$\beta$-stretching of the fragments.
 The new phenomenon could be experimentally observed by the triple gamma
 coincidence technique between the fragments and $^{10}$Be.

\vskip 1.2truecm

  PACS number : 25.85.Ca,27.90.+b

\vfill \eject

\hskip 0.25truecm { \bf 1.  Introduction }

\vskip .75truecm

 The cold (neutronless) fission of
 many actinide nuclei into fragments with masses from 
 $\approx$70 to $\approx$ 160 is nowadays a well studied
 phenomenon [1-8]. Since the final nuclei are generated in
 their ground states or some low excited states,
 these decays were soon related to the
 spontaneous emission of light nuclei (cluster radioactivity)
 such as
 alpha particles and heavier clusters ranging from
 $^{14}$C to $^{34}$Si [9,10].
 All these experimental findings confirmed the theoretical predictions 
 regarding the cold rearrangement processes of large
 groups of nucleons from the ground state of an initial nucleus to 
 the ground states of two or three final fragments [11,12]. 
 Indeed, some fragmentations of heavy nuclei involving
 more than two final fragments have been
 also observed. In the spontaneous and thermal neutron induced fission of
 heavy nuclei the third fragment is usually
 a light charged particle (LCP), the most probable
 being an alpha particle [13-16]. Heavier clusters like
 $^{10}$Be, $^{14}$C, $^{20}$O, $^{24}$Ne, $^{28}$Mg
 and $^{34}$Si [15] have also been detected in these hot 
fragmentations.

   It is very important to establish theoretically and
 experimentally if cold (neutronless) ternary fragmentations
 similar to the cold binary ones are existing in nature.
 This new phenomenon will be equivalent to cluster
 radioactivity during the fission. Such cold ternary 
 decays will produce all three fragments with
 very low or even zero internal excitation energy
 and consequently with very high kinetic energies.
 Their total kinetic energy $TKE = Q_{t} - TXE$  will be close
 to the corresponding ternary decay energy $Q_{t}$.
 In order to achieve such large TKE values, the three final fragments
 should have very compact shapes at the scission point and 
 deformations close to those of their ground states,
 similar to the case of cold binary fragmentations [17,18].

  The first direct observation of cold (neutronless) binary
 fragmentations in the spontaneous fission of $^{252}$Cf was
 made [4,5], by using the multiple Ge-detector Compact Ball
 facility at Oak Ridge National Laboratory, and more recently
 with the Gammasphere consisting of 72 detectors [5,6].

 From these Gammasphere data, it was possible to observe for 
 the first time directly the cold (neutronless) alpha ternary
 fission yields.
 Only the correlations between the two heavier fragments
 were observed unambiguously, by using the triple-gamma
 coincidence technique. 
 Also, in these cold fragmentations, some indications of third light 
fragments as  $^{6}$He, 
 $^{10}$Be and $^{14}$C  clusters were possible [19,20]. 
 Soon direct correlations with the gamma
 rays emitted from LCP may be also possible.
 In this way the accidental coincidences of fission fragments 
 with the binary partners, which represents the main background for
 ternary fragmentations, are eliminated.

 In the present paper, based on a  cluster model, we estimated the 
 relative isotopic yields for the spontaneous cold (neutronless)
 $^{10}$Be -ternary fission of $^{252}$Cf. These isotopic yields
 are given by the ratio of the penetrability through the
 potential barrier between the two final heavier
 fragments for a given mass and charge splitting, over
 the sum of penetrabilities for all possible (neutronless)
 fragmentations. We studied the influence of the fragment excitation
energies on the yields, by including the level densities and the
$\beta$-stretching of the fragments.
The corresponding barriers were evaluated
 using the double folding potential with M3Y nucleon-nucleon
effective interactions and realistic ground state deformations 
including the 
 octupole and hexadecupole ones [6, 21].
 The light cluster was considered as spherical.
 We assumed first that the two heavier fragments penetrate
 the thin barrier existing between them and later-on the LCP
 is jumping over the potential barrier defined by
 the  interaction between the light cluster
 and the two deformed heavier fragments.
 We expect that the fragment mass distribution in cold 
 $^{10}$Be ternary fission of $^{252}$Cf to be similar with that from
 cold binary fission of $^{242}$Pu.

\vskip 1.truecm

\hskip 0.25truecm  { \bf 2. Potential Barriers }

\vskip .75truecm

   In the present paper we consider a cluster model,
 similar to the one-body model used for the description of cluster
 radioactivity [9]. We assume that the initial nucleus
 is already separated into three parts, two heavier ones and
 a cluster, i.e. no preformation factors for the fragments
 are taken into account. An advantage of this model is
 that the barrier between the two heavier fragments and the barrier
 between the light cluster and the heavier fragments
 can be calculated quite accurately due to the fact that the
 touching configurations are situated inside of the barriers.
 The $Q$ values and the deformation parameters contain all
 nuclear shell and pairing effects of the corresponding
 fragments. 

We have calculated the barriers using the double folding model for heavy 
ion interaction 
 $$ V_{F} ({\bf R}) = \int d{\bf r}_{1} d{\bf r}_{2}~
\rho_{1} ({\bf r}_{1}) \rho_{2} ({\bf r}_{2}) v({\bf s}) \eqno(1) $$
where $\rho_{1(2)}({\bf r})$ are the ground state one-body densities 
of the fragments (not necessarily spherical) and $v$ is the $NN$ 
effective interaction. The separation distance between two interacting 
nucleons is denoted by ${\bf s}={\bf r}_{1}+{\bf R}-{\bf r}_{2}$, where
$R$ is the distance between the c.m. of the two fragments.
For simplicity we have choosen the $G$-matrix M3Y effective interaction 
(see [22] for a review) which is representative for the so called local 
and density independent effective interactions. This interaction is 
particularly simple to use in folding models since it is parametrized  
as a sum of 3 Yukawa functions in each spin-isospin $(S,T)$ channel.
Only the isoscalar and isovector components have been retained in the 
present study for the central heavy ion interaction.
The spin-dependent components have been neglected since for a lot of 
fragments involved in the calculation the ground state spins are unknown. 
Moreover, the spin-spin component of the heavy-ion potential is of the 
order $1/A_{1}A_{2}$ and can be safely neglected here.

The M3Y interaction is dominated by the exchange component, therefore it 
is extermely important to include this component in the barrier 
calculation in an accurate way. The most important contribution comes 
from the one-nucleon knock-on exchange term, which leads to a nonlocal 
kernel. The range of the nonlocality behaves as $\mu^{-1}$ , where 
$\mu=A_{1}A_{2}/(A_{1}+A_{2})$ is the reduced mass of the interacting 
system, and therefore the nonlocal potential is reduced in the present 
case to a zero range pseudopotential 
$\hat{J}_{00} \delta({\bf s})$, with a strength depending slightly on the 
energy. We have used the common prescription [22]
$\hat{J}_{00}$ = -276~MeV fm$^{3}$
neglecting completely the small energy dependence. For example, the 
odd-even staggering in the $Q$-value for a fragmentation channel, which 
is tipically of the order $\Delta Q$=2 MeV, leads to a variation with 
$\Delta \hat{J}_{00}$=-0.005$\Delta Q/\mu$ MeV$\cdot$fm$^{3}$ with 
$\mu\approx$100.
The one-body densities in (1) are taken as Fermi distributions in the 
intrinsic frame
$$\rho({\bf r}) =\frac{\rho_0}{1+e^{\frac{r-c}{a}}}\eqno(2)  $$ 
with $c=c_{0}(1+\sum_{\lambda\geq 2}\beta_{\lambda}Y_{\lambda 0}(\Omega))$.
Only  static axial symmetric deformations are considered. The half radius 
$c_{0}$ and the diffusivity are taken either from liquid drop model [27] or
from the spherical HF calculations [23]. The normalization constant $\rho_0$
is determined by requiring the particle number conservation
$$\int r^{2}dr~d\Omega\rho(r,\Omega) = A\eqno(3) $$
and then the multipoles are computed numerically
$$ \rho_{\lambda}(r) = \int d\Omega\rho(r,\Omega)Y_{\lambda 0}(\Omega).
\eqno(4)$$
Once the multipole expansion of the density is obtained, the integral in 
(1) becomes 
$$
V_{F}({\bf R},\omega_{1},\omega_{2}) 
= \sum_{\lambda_{1}\mu_{1}\lambda_{2}\mu_{2}} 
D_{\mu_{1}0}^{\lambda_{1}}(\omega_{1})
D_{\mu_{2}0}^{\lambda_{2}}(\omega_{2})
I_{\lambda_{1}\mu_{1}\lambda_{2}\mu_{2}} \eqno(5)$$
where [24,25]
$$I_{\lambda_{1}\mu_{1}\lambda_{2}\mu_{2}} = 
\sum_{\lambda_{3}\mu_{3}}B_{\lambda_{1}\mu_{1}\lambda_{2}\mu_{2}}^
{\lambda_{3}\mu_{3}}
\int r_{1}^{2}dr_{1} r_{2}^{2}dr_{2} 
\rho_{\lambda_{1}}(r_{1}) \rho_{\lambda_{2}}(r_{2})
F_{\lambda_{1}\lambda_{2}\lambda_{3}}^{v}(r_{1},r_{2},R)
\eqno(6)$$
and 
$$ 
F_{\lambda_{1}\lambda_{2}\lambda_{3}}^{v}(r_{1},r_{2},R) = 
\int q^{2}dq{\tilde v}(q)
j_{\lambda_{1}}(qr_{1})j_{\lambda_{2}}(qr_{2})j_{\lambda_{3}}(qr_{3}).
\eqno(7)$$
Above, $D_{\mu 0}^{\lambda}(\omega)$ stands for Wigner rotation matrix 
describing the orientation $\omega$ of the intrinsic symmetry axis with 
respect to the fixed frame, ${\tilde v}(q)$ denotes the Fourier transform 
of the interaction and $j_{\lambda}$ are spherical Bessel functions. The 
matrix $B$ in (6) is defined in [24] and contains selection rules for 
coupling angular momenta. For example only 
$\lambda_{1}+\lambda_{2}+\lambda_{3}=$even, are allowed.
When $\beta_{\lambda}\neq 0$, $\lambda=2,3,4$ for both fragments, then 
the sum in (5) involves 32 terms for a nose-to-nose configuration and 
$\lambda_{3}\le 6$.   
Special care has been payed to obtain numerically the integrals involved in 
expressions (4) and (6-7). For most of the fragmentation channels studied 
here, large quadrupole, hexadecupole, and ocasionally octupole 
deformations are involved. Therefore a Taylor expansion method for 
obtaining the density multipoles cannot be considered. On the other hand, 
a large quadrupole deformation induces according to (4) nonvanishing 
multipoles with $\lambda$=4 and 6 even if $\beta_4$=$\beta_6$=0. 
Therefore for a correct calculation of 
(4), a numerical method with a truncation error of order O$(h^7)$ is 
needed in order to ensure the orthogonality of spherical harmonics with 
$\lambda\le 6$.

Performing the integrals (6) and (7) we have used a numerical method with 
a truncation error of the order O$(h^9)$. All short range wavelength 
($q\le 10$~fm$^{-1}$)have been included and particular care has been taken 
to ensure the convergence of the integrals with respect to the 
integration step and the range of integration.

The asymptotic part of the barrier is determined essentially by the 
Coulomb multipoles which are obtained also as double folding integrals 
involving charge densities. 
 For $R>>r_{1}+r_{2}$, the Coulomb kernel in (7) 
behaves as [24]
$$F_{\lambda_{1}\lambda_{2}\lambda_{3}}^{C}(r_{1},r_{2},R) =
2\pi^{2}
\frac{(2\lambda_{3}+1)!!}{(2\lambda_{3}+1)(2\lambda_{1}+1)!!(2\lambda_{2}+1)!!}
\frac{r_{1}^{\lambda_{1}}r_{2}^{\lambda_{2}}}{R^{\lambda_{3}+1}}
\delta_{\lambda_{3},\lambda_{1}+\lambda_{2}}.
\eqno(8)$$
If we introduce the moments of the charge density as 
$$
Q_{\lambda}=\sqrt{\frac{4\pi}{2\lambda+1}}
\int_{0}^{\infty} r^{2} dr \rho_{\lambda}(r) r^{\lambda}
\eqno(9)$$
where $Q_{0}=Z$ (atomic number)
then the $\lambda_{3}=2$ component of function (6) behaves for 
$R\rightarrow\infty$ as 
$$
(C_{000}^{202})^{2}\frac{Z^{1}Q_{2}^{2}+Z^{2}Q_{2}^{1}}{R^3}.
\eqno(10)$$

   At the scission configuration we assumed two coaxial 
deformed fragments in contact at their tips.
  For quadrupole deformations we choose two coaxial prolate spheroids 
due to the fact that the prolate shapes are favoured in fission. 
It is known that for each oblate minimum always corresponds another 
prolate minimum. For pear shapes, i.e. fragments with quadrupole and 
octupole deformations, we choose opposite signs for the octupole 
deformations, i.e. nose-to-nose configurations (see Fig.1). 
For hexadecupole deformations we choose only positive 
signature, because it leads to a lowering of the barriers in comparison 
with negative ones and consequently they are much more favoured in 
fission (see Fig.2).   

 In order to ilustrate the influence of deformations on the barriers 
we displayed in Fig.3  the M3Y-folding 
multipoles for $^{98}$Sr and $^{144}$Ba with all deformations included. 
The octupole component is large in the interior but gives negligible
contribution in the barrier region in contrast with the 
hexadecupole one.
Next, in Fig.4 we are illustrating for the same partners the
cumulative effect of high rank multipoles on the barrier. We stress
the correct asymptotic behaviour of multipoles, especially of the 
component $\lambda_{3}=2$ of function (6) given by expression (10) which 
survives up to very large distances if the quadrupole deformation is 
large.

\vskip 1.5truecm
\newpage
 
\hskip 0.25truecm  { \bf 3. Isotopic Yields with Liquid Drop Parameters}
 
\vskip .75truecm

   We should like to stress again that in our simple cluster
 model we neglect the preformation factors for different 
 channels, i.e. we use the same frequency factor $\nu$  for the
 collisions with the fission barrier for all fragmentations.
 It is generally known that
 the general trends in alpha decay of heavy nuclei
 are very well described by barrier penetrabilities, the preformation
 factors becoming increasingly important only in the vicinity of
 the double magic nucleus $^{208}$Pb. On the other hand
 the cold binary fragmentation of $^{252}$Cf was also reasonably
 well described using constant preformation factors [6,28]. Similarly
 we expect that the ternary cold splittings could be described in the
 first order approximation only by the barrier penetrabilities.
 Eventually, as the experimental data become more accurate we would
 be able to extract some fragment preformation factors
 and discuss the related nuclear structure effects.
 Presently, it is too early to compute cluster preformation factors
 on the nuclear surface of a fissioning nucleus [29].
 
  In the laboratory frame of reference the $z$-axis was taken as
 the initial fissioning axis of the two heavier fragments,
 with the origin at their point of contact. We assumed that
 the three bodies are moving in the $(z, x)$ plane.
 The potential barriers $V_{HL} - Q_{HL}$ between the two
 fragments are high but rather thin with a width
 of about 2 to 3 fm. As an illustration, we show in Fig.5
 a typical barrier between $^{142}$Xe and $^{100}$Zr,
 as a function of the distance $R_{HL}$ between their 
 center of mass. Here $Q_{HL} = Q_{t} - Q_{c}$ is the decay
 energy for binary fragmentation (in our case that of $^{242}$Pu)
 and $Q_{c} =$ 8.71 MeV is the $^{10}$Be decay energy from $^{252}$Cf.

   On the other hand the LCP is initially situated in the 
 potential well which is created by the sum of the potentials between 
 the LCP and the two heavier fragments. As an
 illustration we present in Fig.6 this potential barrier for 
 the light cluster, as a function of its position in the ($z, x$) plane,
 at three different values of the inter-fragment distance $R_{HL}$.
 The corresponding ternary splitting is the same as in Fig. 5.
 We can see that as the distance between the two heavier
 fragments increases the LCP potential well is narrowing
 and its bottom rises, forcing the cluster to jump
 over the barrier and to be repelled along the $x$-axis
 by the Coulomb field of the other two fragments.
 For the two fragments, the exit point from
 their potential barrier is at $R_{HL}$ typically
 between 15 and 16 fm (see Fig.5) which supports our
 cluster model. Evidently from the top of the cluster
 barrier we can compute the classical trajectories of
 the three fragments as a function of time.

Due to the fact that the barrier between the two
 heavier fragments is much thinner than the barrier between
 the LCP and the heavier fragments,
 in our model first the two heavier fragments
 penetrate the potential barrier between them and later-on 
 the LCP is emitted. In such a model the mass distributions
 of the heavier fragments are not influenced by the cluster
 trajectories. Consequently these mass distributions are very similar
 to that of the cold binary fission of an initial nucleus
 leading to the same heavier fragments, i.e. in our case
 $^{242}$Pu.
 This mechanism is supported by the comparison between the
 experimental data concerning the fission mass distributions
 in binary and alpha-accompanied fission of $^{235}$U [30].
 From these data the experimentalists
  concluded that the LCP is preferentially emitted
  by the light fragment [14]. We should like to mention that
 a sequential emission of a $^{10}$Be cluster from
 the already separated fragments is not possible due to the
 fact that these are very neutron-rich nuclei with negative
 or close to zero
 $Q$ values. In addition, the presence of a Coulomb barrier
 further hinders LCP emissions from the heavier fragments in comparison
 with the neutron evaporation process at excitation energies
 above 6-7 MeV. On the other hand, it is known that for
 the mass distributions in asymmetric spontaneous fission of
 the lighter actinides compared to the heavier ones, the position of
 the heavy mass peak remains unchanged  
 while the light mass peak moves to lower $A_{L}$ values [14,30].
 Thus from our model
  we conclude that the mass distribution of fission fragments
 in cold ternary fission is almost identical with the mass
 distribution for the cold binary fission of the
 daughter nucleus (i.e. the initial nucleus from which the LCP
 was extracted). This looks like the LCP was emitted from
 the light fragment.

   The penetrabilities through the 
 double-folded potential barrier between the two heavier
 fragments were calculated by using the WKB approximation 

  $$P = \exp \left\lbrace -{2 \over \hbar} \int_{s_{i}}^{s_{o}} 
  \sqrt{~ 2 \mu~ [~V_{HL}(s)-Q_{HL}~]~}~~ds \right\rbrace   
\eqno(11) $$
  where $s$ is the relative distance,
 $\mu$ is the reduced mass and $s_{i}$ and $s_{o}$ 
 are the inner and outer turning points, defined
 by $V_{HL}( s_{i} ) = V_{HL}( s_{o} ) = Q_{HL} $.

The barriers were computed with the LDM parameters $a_p=a_n$=0.5~fm, 
$r_{0p}=r_{0n}=(R-1/R)A^{-1/3}~$fm with 
$R=1.28A^{1/3}+0.8A^{-1/3}-0.76$.

   Accurate knowledge of $Q$ values is crucial for the calculation,
 since the WKB penetrabilities are very sensitive to them.
 We obtained the $Q$ values from experimental mass tables 
 [26], and for only a few of the fragmentations the nuclear masses
 were taken from the extended tables of M\"oller et al. [27]
 computed using a macroscopic-microscopic model.

 In this paragraph we consider only the relative isotopic yields 
 corresponding to true 
 cold (neutronless) ternary fragmentations in which
 all final nuclei are left in their ground state.
 These ternary relative isotopic yields are given by the expression
($A_{1}=A_{L},A_{2}=A_{H}$)
 $$ Y ( A_{1}, Z_{1} ) = { P ( A_{1}, Z_{1} ) \over \sum_{A_{1} Z_{1}}
  P ( A_{1}, Z_{1} ) } ~~\cdot \eqno(12)  $$
As fragment deformations we choose the ground state deformations of 
M\"oller et al.[27] computed in the frame of the macroscopic-microscopic
model. Due to the fact that the influence of fragment deformations on
barrier penetrabilities, i.e. on ternary yields, is extermely large we
represented in Fig.7 these deformations for the light $A_{1}$ and
heavy $A_{2}$ fragments, separately for odd and even charge $Z$. We
can see that the light fragments , have mainly quadrupole deformations
in contrast with the heavy fragments, which have all
deformations. 
The octupole deformations are existing for a small heavy fragment mass 
number region 
141 $\leq A_{2}\leq 148$.
The fragments with mass number $A_{L}\leq 92$
and $A_{H}\leq 138$  are practically spherical.

The computed M3Y-fission barriers heights, for different assumptions: no 
deformations,
including the quadrupole ones, including the quadrupole and octupole
ones and for all deformations, together with the corresponding 
$Q$-values 
are represented in Fig.8 for odd $Z$ and even $Z$ separetely. We can
see that the largest influence is due to the quadrupole deformations
but also the hexadecupole ones are lowering the barriers very
much. 
A strong lowering of barriers height starts around $A_{1}=95$, 
$A_{2}=139$ and finish at $Z_{1}=41$, $A_{1}=104$ and 
$Z_{1}=42$, $A_{1}=103$ which corresponds to the sudden increase of
deformations for $Z_{2}=53$, $A_{2}=138$ and $Z_{2}=52$, $A_{2}=139$
(Fig.7).
The octupole deformations in the mass region  
$141\leq A_{2}\leq 148$ have a smaller
effect as we expect. This is a illustration of the difference between
cluster radioactivity, which is due only to the large $Q$-values and
the cold fission which is due mainly to the lowering of the barriers
due to the fragment deformations. Both processes are cold
fragmentation phenomena.

The computed yields in percents, for the splittings represented by
their fragment deformation parameters in Fig.7, or by their barrier 
heights in Fig.8 are given in Fig.9 for
spherical fragments ($\beta=0$), for quadrupole deformations ($\beta_{2}$)
and for all deformations ($\beta_{2}+\beta_{3}+\beta_{4}$). We can see
that for spherical fragments the splittings with the highest
$Q$-values, which correspond to real spherical heavy fragments(see
Fig.7), i.e. for charge combinations $Z_1/Z_2=$ 44/50, 43/51 and 42/52 
are the predominat ones. As we mentioned before this situation is similar 
with the cluster radioactivity were the main fact is the $Q$-value. Due to
the staggering of $Q$-values(see Fig.8) the highest yields are for even-even
splittings. 
By including the $\beta_2$ deformations few asymmetric splittings exists.
For all deformations more asymmetric yields appear. Now the principal 
yields are for $Z_{1}/Z_{2}$= 38/56 and 40/54.
This is due to the fact that the influence of the fragment 
deformations on the yields could compensate the influence of
$Q$-values. This illustrate the fact that cold fission is a cold
rearrangement process in which all deformations are playing the main
role and not the  $Q$-values. The staggering for odd $Z$ fragmentations 
like $Z_{1}/Z_{2}$ = 39/55, 41/53 and 43/51 or odd $N$ fragmentations 
is recognized at first glance. We shall see later on that by the 
introduction of the density levels this staggering is reversed. The 
largest yields will be for odd $Z$ and/or $N$ fragmentations.

In the next figures, we represented the mass yields 
$Y_{A_{1}}=\sum_{Z_{1}}Y(A_{1},Z_{1})$ (Fig.10) and the charge yields 
$Y_{Z_{1}}=\sum_{A_{1}}Y(A_{1},Z_{1})$ (Fig.11)
for spherical fragments ($\beta_i$=0), for quadrupole ($\beta_2\neq$0) ones 
and for all deformations ($\beta_i\neq$0). We can see that for spherical 
fragments the main mass yields are $A_{1}$=108, 110 and 112 for 
spherical or for prolate shapes. The hexadecupole deformations shifts the main yields 
to $A_{1}$=96-104. The experiment will decide if these deformations are 
participating to the cold fission process.
We stress the odd-even mass and charge staggering which already existed at 
individual yields $Y(A_{1},Z_{1})$.

 These predictions are
 very useful as a first guide for unfolding
 the cold ternary yields from the very complex experimental
 gamma-ray spectra containing the contributions from over 100
 fission fragments. We should like to stress the correlation
 between the ternary fission $Q_{t}$ values and the isotopic
 yields. Usually for a given mass fragmentation the highest
 yield corresponds to the charge splitting with the highest
 $Q_{t}$ value, but in a few cases 
 where the fragment deformations are large,
 this correspondence is reversed. This support our interpretation
 of cold fission as cluster radioactivity.

\vskip 1.truecm

\hskip 0.25truecm { \bf 4. Isotopic yields with Hartree-Fock parameters}
\vskip .75truecm

In order to obtain the parameters of the one-body densities involved in 
the folding integral, we have performed a large scale (162 nuclei) 
standard spherical Hartree-Fock(HF) calculation using the energy density 
formalism of Beiner and Lombard [23]. 
The functional $F_{1}$ was used for all nuclei since it reproduces 
correctly the bulk properties of spherical nuclei (binding, separation 
energies, single particle levels, charge radii, etc.).
The shell model occupation probabilities were used in constructing the 
single particle orbitals. Pairing interactions were neglected. This is a
limitation in our calculation. However this type of correlations are 
implicitely included in our densities since we used the deformations 
given by the macroscopic-microscopic model of M\"oller et al.[27].
For each nucleus considered, the HF density was fitted with a Fermi 
distribution in the range 2-18 fm. We obtained  good quality fits in the 
surface region which largely determines the diffusivity. The fitting 
parameters (reduced radii and diffusivities) were displayed in Fig.12 which
clearly emphasize the effect of partial filling of nuclear subshells.
In Fig.13 we represented the barrier heights for different assumptions 
on fragment deformations : spherical, prolate, pear-shape and with 
hexadecupole as a function of the light fragment mass.
We see the same strong dependence of barriers heights on the 
fragment deformations like in 
the case of liquid drop model parameters(Fig.9). In Fig.14 we give the cold 
fission yields in percents for zero fragment deformations, with quadrupole 
and 
with all deformations included. 
In order to understand the beahviour of HF-parameters we represented in 
Fig.15: the inner turning point $(r_1)$ the outer turning point $(r_2)$, 
the radius at the barrier height $(r_b)$ and the radius of the touching
configurations $r_t=(r_{A_1}^2)^{1/2}+(r_{A_2}^2)^{1/2}$ as a function of 
the light fragment mass.
We stress that $r_t$ is a smooth function of $A_1$, while
$r_1$ and $r_b$ are strongly influenced by deformations and $r_2$
reflects the odd-even staggering of $Q$-values.

 In Fig.16 we represented the mass yields 
$Y_{A}=\sum_Z{Y(Z,A)}$ and in Fig.17 the charge yields. The general 
behaviour is similar with the trend given by the yields computed with 
the liquid drop model parameters (see Figs.10,11). 
The geometrical parameters of the one-body densities (reduced radius and 
diffusivity), provided they have realistic values, influences to a lesser 
extent the relative yields, as compared with $Q$-values and deformation
parameters.
\vskip 1.truecm

 \hskip .25truecm { \bf 5. Ternary yields as Function of Fragment 
Excitation Energy}
\vskip .75truecm

In the following we shall study the influence of the level density and of 
the total excitation energy ($TXE$) of the fragments on ternary yields. 

Following a suggestion of Schwab et al. [3] we define the level 
density for the composite system as a function of total excitation energy 
($TXE$)  
 $$\rho(TXE) = \int_{0}^{TXE}\rho_{L}(e)\rho_{H}(TXE-e)de ~~\cdot 
\eqno(13)  $$
where $\rho_{L(H)}$ are the individual level densities of the fragments.
The above definition is consistent with the hypothesis of a uniform 
distribution of the total available excitation energy between the fragments.
For excitation energies larger than 1MeV we have used the usual formula 
obtained within the Back Shifted Fermi Gas Model(BSFGM). 

The level density parameter and the fictious ground state position 
($a$ and $\Delta$ respectively) were taken form the global analysis 
of Dilg et al.[31]. 
For smaller excitation energies the BSFGM level density was smoothly 
joined with a formula by Grossjean and Feldmayer [32] which avoids a 
singularity close to $e\rightarrow 0$. 
We should like to mention that the introduction of a fictious ground 
state position, according to odd -even $Z$ or $N$, change completely the 
odd-even effect in $Z$ or $N$. If before the even-even splittings 
were favoured, now the largest yields are for odd-odd $Z$ or/and $N$ 
splittings in agreement with the experimental data.

Second, we shall consider the change of deformation due to the fragment 
excitation energy.
Like in our previous papers [28,33] we assume that the total excitation 
energy $TXE=Q-TKE$ will lead to a supplementary deformation of the 
fragments viewed as a $\beta$-stretching along the elongation 
axis. According to this model at the scission point the fragments will 
have slightly bigger defformations than the ground state values. The 
excitation energy $TXE$ is supposed to be divided proportionaly to the 
mass of each fragment, i.e. $E_i^*=\frac{A_i}{A_1+A_2}TXE~(i=1,2)$. Then 
the $induced$ deformation coming-up from the $\beta$-polarization of fragment
$i$ is given by the expression [28]:
 $$ \beta(E^*)=\left\lbrace \beta_0^2+\frac{\hbar}{2B\omega_\beta}
\left ( {2E^*\over \hbar\omega_\beta}+1
\right )\right\rbrace^{1/2}
\eqno(14) $$
where $\beta_0$ is the ground state deformation of the corresponding 
fragment, B is the mass parameter and $\omega_\beta$ is the 
frequency for the $\beta$-vibrations, both being evaluated from the 
experimental data. The use of the $\beta$-stretching is motivated by the 
fact that the cold fission yields are increasing with the excitation 
energy [1,3,8]. The level density is also increasing, but not enough to 
compensate the decrease of the barrier penetrabilities with the increase 
of the excitation energy. Consequently by $\beta$-stretching we increase 
the deformations, i.e. we decrease the barriers which leads to the 
increase of the barrier penetrabilities which allows to reproduce such a 
trend of cold fission yields. In a previous paper [33] we succeded to 
describe excellently the behaviour of the experimental cold fission 
yields for the thermal neutron fission of $^{233}$U as a function of the 
excitation energy.

We would like to stress that according to expression (14) the excitation 
energy will enhance the deformations especially for fragments with large 
deformations. On the other hand in cold fission the two fragments must 
have a stretched out scission configuration in order to have a large 
penetrability, i.e. large yield. Consequently only the combinations  
for which the increase of deformations is large enough are 
observed experimentally. This may change the order of fragmentation 
channels which are mainly contributing to a given isotopic yield.

The penetrabilities computed with $\beta$-stretching multiplied with the 
level density for excitation energies $E^*$=1, 3 and 5 MeV are 
represented in Fig.18. We 
can see that more combinations are contributing to the cold fission 
yields with the increase of the excitation energy. Only a detailed 
comparison with the experimental data will give an answer if the present 
dynamical model describe correctly the experimental data. Obviously the 
level densities fitted for the outer region with neutron rich fragments and 
the introduction of other types of polarizations are necessary.   
\vskip 1.truecm

 \hskip .25truecm { \bf 6. Experiments}

\vskip .75truecm

 Up to now, in order to study the binary and alpha ternary fission of 
$^{252}$Cf, a source of strength 
 $\sim 6 \times  10 ^{4}$ fissions/s, sandwiched between two
 Ni foils of thickness 0.5 mil and then sandwiched between 2 mil
 thick Al foils, was placed at the center of the
 Gammasphere with 72 Compton suppressed
 Ge detectors. A total of 9.8x10$^{9}$
 triple or higher fold coincidence events were recorded.
 A $\gamma$ - $\gamma$ - $\gamma$ "coincidence cube" 
  was built using the RADWARE software program [34].
 In the fission of $^{252}$Cf, about 100 different final fragments 
 are produced. 
First, these primary fragments emit several neutrons until the excitation
 energy of the fragment is below the neutron binding 
 energy ($\sim$ 6 MeV). The excited primary 
 fragments are too neutron-rich to emit charged particles such as
 protons, alpha particles or light charge particles as $^{10}$Be. 
Then, the secondary fragments decay to their ground states by
 the emission of $\gamma$ -rays. 
Only the correlations 
between the two heavier fragments were observed unambigously by using the 
triple-gamma coincidence. 

 The neutron multiplicities and 
 the correlated yields of $Z$ secondary binary fragments were
 determined by setting a double gate on the light fragment
 and measuring the gamma intensities in the heavy fragment
 with different number of evaporated neutrons.
 For the odd-odd fragmentations we considered the total intensities
 obtained from summing all the gamma transitions to the ground
 state. Correcting the number of counts for
 the detector efficiency and for internal conversion,
 we obtain the gamma transition 
 relative yields for the different neutron channels.
 Then the sum of these ground state transition yields are
 normalized, for a given light fragment, to the Wahl Tables which give
 the estimated total
 isotopic yields in the binary fission of $^{252}$Cf [35].
 If one isotope from the heavy fragment is missing we evaluate
 its corresponding yield by interpolation from its neighbors
 with a Gaussian. A cross-check is necessary by imposing
 a double gate on the heavy fragment and determining the gamma
 intensities of the corresponding correlated light fragments.
 Again the sum of these yields are normalized to the Wahl
 Tables [35] for the heavy fragment. The final isotopic yields
 must be consistent.
In cases where either the background is large
or when the gates are complex because of multiple $\gamma$-rays of 
the same energies a third determination was made by setting a gate
 on the light fragment and another gate on the heavy fragment.
 Determining the intensities in both fragments and knowing
 the branching ratios between different transitions we can
 determine again the binary yields [5,6]. We should like to mention
 that presently the spectra of odd-$Z$ nuclei are not known,
 which does not allow us to determine experimentally the odd-$Z$
 isotopic yields.

  The cold (neutronless) binary yield represents only 0.2$\%$ 
from the total fission yield. Due to the fact that the values of 
the alpha ternary yields are comparable with the values of cold
binary yields the background must be similar [36].

The $^{10}$Be-ternary yields are at least 100 times smaller than 
$\alpha$-ternary yields, i.e. 0.002$\%$ [37].
For the corresponding cold yields we expect the same ratio.
Consequently we have to be very careful with the accidental 
coincidence of the two final fragments with their binary partners.
The detection of fragments by particle detectors in coincidence with the 
$\gamma$-rays of the fragments does not solve this problem. The binary 
partners of the fragments
will be present by their characteristic $\gamma$-rays. 
Any how the large background will be reduced. At least, due to the fact 
that we could detect the $TKE$ of the fragments, we can
determine the yields as function of fragment $TKE$. Another possibility
is to detect directly the $^{10}$Be-nucleus by a particle detector.
Unfortunately the large background of alpha particles will cover such a 
small yield. A much more reasonable possibility look like to use a 
characteristic $\gamma$-ray of $^{10}$Be in coincidence with the fragments
of $\gamma$-rays. This will eliminate the large background due to the
binary partners of the fragments.

  We should stress here that up to now the experimentally determined isotopic
 yields by triple $\gamma$-coincidence method are integrated yields. In the 
spontaneous fission experiment
 of $^{252}$Cf the majority of the binary and ternary splittings lead
 to highly excited final nuclei which after neutron evaporation are
 decaying to the lowest states by gamma cascades. Less frequently,
 there are cold fragmentations which leave the final nuclei in
 their ground or first excited states. We define these 
 cold fission experimental yields as integrated yields since they 
 collect the contributions of all (neutronless) transitions over
 a whole range of TXE from zero up to at least the neutron
 binding energy, from where the evaporation of a first neutron
 becomes possible. 
For a deeper interpretation we need the yields as a function of the
excitation energy of fragments. 

\vskip 1.truecm

\hskip .75truecm { \bf 7. Discussions and Conclusions }

\vskip 1.truecm

   The cold alpha ternary fission of a heavy nucleus ($^{252}$Cf) was
 experimentally recently observed directly for the first time [36].
 Some indications of $^{10}$Be and $^{14}$C ternary fragmentations were
 also obtained [19,20].
 The cold ternary fission events are characterized by very low 
 TXE of the final fragments and high TKE tending to the $Q_{t}$
 value associated to those splittings. Thus the configuration at
 the scission point should be described in these cases by very
 compact shapes, the deformed fragments in their ground states.
 It was already shown that for
 cold binary fragmentations, the ground state deformations
 are a key ingredient for the correct prediction of the most
 favoured splittings and of the isotopic yields [6,17,18,28].
 The cluster model which we used in this
 paper for calculating the isotopic yields 
 associated with $^{10}$Be-accompanied cold ternary
 fission, also predicts a large number of favored ternary
 splittings in which one or both heavier fragments are well
 deformed in their ground states.

   The determination of the scission point configurations in
 the fission of heavy nuclei starting from the experimental
 kinetic energy and angular distributions of the LCP emitted in
 ternary fission, has been a great hope for many years. 
 Unfortunately too many unknown parameters are associated with
 the initial scission point configurations in the case of the
 usual "hot" ternary fission. But for cold ternary 
 fission the initial scission configurations are known : the 
 fragment deformations should be essentially that of the
 ground state deformations. Of course the initial position and
 velocity distributions of the LCP have to be determined from
 their final kinetic energy and angular distributions.

  Presently the spontaneous cold fission of three nuclides,
 namely $^{252}$Cf, $^{248}$Cm and $^{242}$Pu, is under study
 using the triple gamma coincidence technique. The same set of 
 deformation parameters should explain the cold binary fission
 yields in all three cases so that we expect to extract
 new experimental information over different nuclear deformation 
 regions. In addition the cold alpha ternary fission yields
 of $^{252}$Cf should be similar to the cold binary fission
 yields of $^{248}$Cm [21] and the cold $^{10}$Be- ternary fission
 yields of the same parent nucleus should be similar
 to the cold binary fission yields of $^{242}$Pu [38,39].
 Consequently many cross-checks are possible. In addition the
 kinetic energies and angular distributions of light clusters
 emitted in cold ternary fission will provide new important
 insight on the fragmentation processes of heavy nuclei.

\vskip 0.5truecm
\centerline { Acknowledgements.}

 The work at Vanderbilt University was supported in part by the U.S.
 Department of Energy under grant No. DE-FG05-88ER40407. The Joint 
 Institute for Heavy Ion Research has member institutions the University
 of Tennessee, Vanderbilt University and Oak Ridge National Laboratory.
 It is supported by its members and by the U.S. Department of Energy
 through contract No. DE-FG05-87ER40361 with the University of Tennessee.
 A.S., F.C., \c S.M. would
 like to acknowledge the hospitality of Gesselschaft f\"ur
 Schwerionenforschung, Darmstadt, Germany, and A.S. and A.F.  
 of Vanderbilt University, Nashville, U.S.A. during the 
completion of this work.
 We would like to acknowledge the grant received under Twinning Program
 from National Research Council.

\vfill\eject

\vfill \eject

\vskip 7truecm
\centerline {\bf Figure  Captions }
\vskip 1truecm

 ${\bf Fig.~1.}$ Density plots of $^{98}$Sr
and $^{144}$Ba fragments, placed at $R$=15fm, considered with quadrupole and 
octupole deformations. In the upper part are represented the prolate-prolate
, oblate-prolate positions and in the lower part two pear shapes 
nose to back and nose to nose. The positions are given by the deformation 
signs.
 \vskip 1.0truecm

 ${\bf Fig.~2.}$ Same as for Fig.1. The influence of different signs of 
hexadecupole deformations on $^{98}$Sr and $^{144}$Ba densities in the 
presence of large quadrupole and octupole deformations. The penetrability 
is maximized for $\beta_{4}>$0 configurations.

\vskip 1.0truecm

 ${\bf Fig.~3.}$ The influence of the M3Y-folding multipoles on the 
barrier between $^{98}$Sr and $^{144}$Ba. Notice that the main effect 
is due to $\lambda_{3}=2$. The influence of $\lambda_{3}=3$ is large 
but less important in the barrier region compared with the induced 
deformations $\lambda_{3}=5$ and $\lambda_{3}=6$ 
\vskip 1.0truecm

${\bf Fig.~4.}$ The cumulative effect of high rank multipoles on the 
barrier between $^{98}$Sr and $^{144}$Ba. We considered the  
deformations $\beta_{3}$ and $\beta_{4}$ 
much larger than the real ones in order to 
illustrate the effect of deformations.
\vskip 1.0truecm

${\bf Fig.~5.}$ The barrier between $^{142}$Xe and $^{100}$Zr as a function
of the distance $R_{HL}$ between their centers of mass. 
By $Q_{HL}=Q_{t} - Q_{c}$ we 
denote the decay energy of daughter nucleus ($^{242}$Pu)
where $Q_{t}$ is the ternary decay energy for $^{10}$Be ternary cold
splitting and $Q_{c}$=8.71MeV is the $^{10}$Be decay energy from 
$^{252}$Cf.
\vskip 1.0truecm

${\bf Fig.~6.}$ The barrier between the $^{10}$Be cluster and the two 
heavier fragments in the $(z,x)$ plane for three fixed distances 
between them : 
(a) $R_{HL}$ = 12.6, (b) $R_{HL}$ = 16.6 
and (c) $R_{HL}$ = 20.6. Note the larger cluster barrier widths.

\vskip 1.0truecm
${\bf Fig.~7.}$ The assumed $\beta_{2}$, $\beta_{3}$, $\beta_{4}$
ground state fragment deformations [27]. We can see that the light fragments 
$(Z_1,A_1)$ have mainly quadrupole deformations in contrast with the 
heavy fragments $(Z_2,A_2)$. The octupole deformations are existing in a 
small mass region 141$\leq A_2\leq$148. The fragments with masses 
$A_1\leq$90 and $A_2\leq$138 are practically spherical.
\vskip 1.0truecm

${\bf Fig.~8.}$ The barrier heights for all considered fragmentations  
channels represented 
for different charges $Z_1$ and mass numbers $A_1$ of the light fragment.
The strong lowering of the barrier heights at 
$Z_1$=41, $A_1$=104 and $Z_1$=42, $A_1$=103 corresponds to the 
sudden increase of deformations at $Z_2$=53, $A_2$=138 and 
$Z_2$=52, $A_2$=139 (Fig.7). The decrease of barrier heights starts 
approximately at $A_1$=95, respectively at $A_2$=139. 
$Q$ values are represented by slightly larger symbols.    
\vskip 1.0truecm

${\bf Fig.~9.}$ The true cold fission yields in percents for all 
fragmentations channels considered in Figs.7 and 8 computed with the LDM 
parameters, for spherical nuclei, with the inclusion of quadrupole 
deformations and with all deformations.
In the bottom histogram, all yields $\leq$ 10$^{-4}$ were set arbitrarily to 
1.5$\cdot 10^{-4}$ to make easy the identification of the calculated 
fragmentation channels.  

\vskip 1.0truecm

${\bf Fig.~10.}$ The mass yields $Y_{A_1}=\sum_{Z_1}Y(A_1,Z_1)$ in 
percents, as a function of light fragment mass computed with LDM parameters.
The same conclusions are obtained as before for separate charge and mass 
splittings. Calculations without deformations ($\beta_{2,3,4}$=0) enhance 
only the spherical region $ A_1 \geq$ 106; the inclusion of 
quadrupole deformations ($\beta_2\neq$0) enhances also the yields with 
$A_1 \geq$96; for all deformations the main mass yields region becomes 
96$\leq A_1 \leq$104.

\vskip 1.0truecm

${\bf Fig.~11.}$ 
The charge yields $Y_{Z_1}=\sum_{A_1}Y(A_1,Z_1)$, in percent, as 
function of light fragment mass computed with LDM-parameters. 
The same conclusions are obtaineed as before for separate charge and 
mass splittings. The calculations with no deformations ($\beta_i$=0)
enhances only the yields with $Z_1 \geq$42 and 44. The quadrupole 
deformations enhance also $Z_1$=38 and 40. The inclusion of all 
deformations enhances mainly the yields with  $Z_1$=38 and 40.      

\vskip 1.0truecm

${\bf Fig.~12.}$ The geometrical parameters of the HF one-body densities, 
i.e. the reduced radii $(r_n,r_p)$ and the diffusivities $(a_n,a_p)$,
for 162 light or heavy fragments included in the splittings represented 
in Figs.7 and 8. The parameters were obtained by fitting the HF one-body  
densities with Fermi density distributions in the range 2-18 fm. 
The effect of partial filling of nuclear subshells is clearly seen.

\vskip 1.0truecm

${\bf Fig.~13.}$ The barrier heights computed with the HF parameters
 for Fermi one-body densities represented separatelly for odd $Z$ and 
even $Z$. Succesively we give the barrier heights for spherical 
fragments, for quadrupole, octupole and hexadecupole deformations 
together with the $Q$-values for the splittings considered in Figs.7 and 
8. The lowest barriers are for 95$\leq A_1 \leq$103 ($Z$-odd) or 
104 ($Z$-even)

\vskip 1.0truecm

${\bf Fig.~14.}$  The ternary yields in percents for the true cold 
$^{10}$Be-ternary splittings computed with the HF parameters for 
different deformation sets : spherical ($\beta_i$ = 0), prolate 
($\beta_2\neq$0) and all deformations ($\beta_2+\beta_3+\beta_4)$. The 
results are similar with LDM-ternary yields represented in Fig.9.
In the bottom histogram, all yields $\leq$ 10$^{-4}$ were set arbitraily to
1.5$\cdot 10^{-4}$ to make easy the identification of the calculated 
fragmentation channels.

\vskip 1.0truecm

${\bf Fig.~15.}$ The inner turning point $(r_1)$, the outer turning 
point $(r_2)$, the radius of the HF barrier heights $(r_s)$ and the 
touching radius $r_t=(r_{A_1}^2)^{1/2}+(r_{A_2}^2)^{1/2}$ as function of 
light fragment mass. We denoted by open symbols odd $Z$ fragments and by
full symbols even $Z$ fragments. The lines connect the symbols with the 
same $Z$. We can see that $r_t$ is a smooth function of $A_1$, that 
$r_1$ and $r_b$ are strongly influenced by deformations and $r_2$ 
reflects the odd-even staggering of $Q$-values.      

\vskip 1.0truecm
 
${\bf Fig.~16.}$ The mass yields in percents 
$Y_{A_1}=\sum_{Z_1}Y(A_1,Z_1)$ computed with the HF one-body density 
parameters for different assumptions concerning the fragment deformations:
spherical fragments ($\beta_i=0$), quadrupole fragments ($\beta_2\neq2$), 
and all deformations ($\beta_2+\beta_3+\beta_4$). Similar with mass 
yields for LDM parameters we obtained for spherical fragments only for the
spherical region $A_1\geq 106$, for prolate fragments additional lighter 
yields for $A_1\geq 96$ and for all deformations the mass yields are
concentrated in the region 96$\leq A_1\leq $104
\vskip 1.0truecm

${\bf Fig.~17.}$
 The charge yields in percents $Y_{Z_1}=\sum_{A_1}Y(A_1,Z_1)$ computed 
with HF one-body density parameters as a function of the light fragment 
mass for different assumptions concerning the fragment deformations. For
spherical fragments we have charge yields only for $Z_1$=42 and 44.
For prolate fragments the mass yields start with $Z_1$=38 and 40. For all 
deformations the main charge yields are for $Z_1$=38 and 40.
\vskip 1.0truecm

${\bf Fig.~18.}$ The cold fission yields at excitation energies 
$E^*$=1,3 and 5 MeV, i.e. with modified penetrabilities due to the 
$\beta$-stretching and multiplied by level densities for all fragmentation
channels in Figs.7 and 8. In the bottom histogram, all yields $\leq 10^{-13}$
were set arbitrarely to 1.5$\cdot 10^{-13}$ to make easy the 
identification of the calculated yields. We can see that fission yields 
are increasing with the excitation energy and that odd $Z$ or/and odd $N$
splittings are larger than the even ones.

\vfill \eject

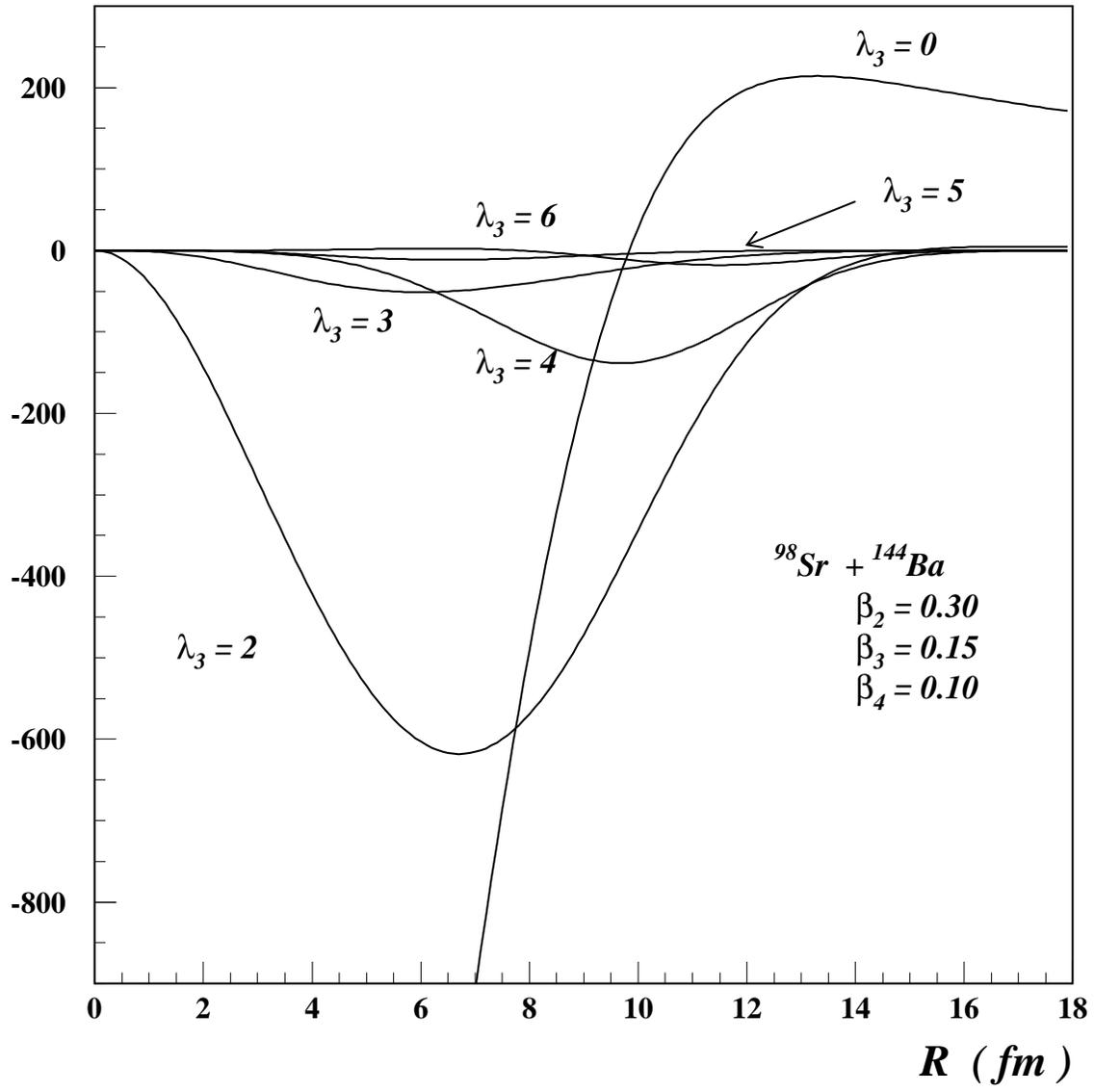
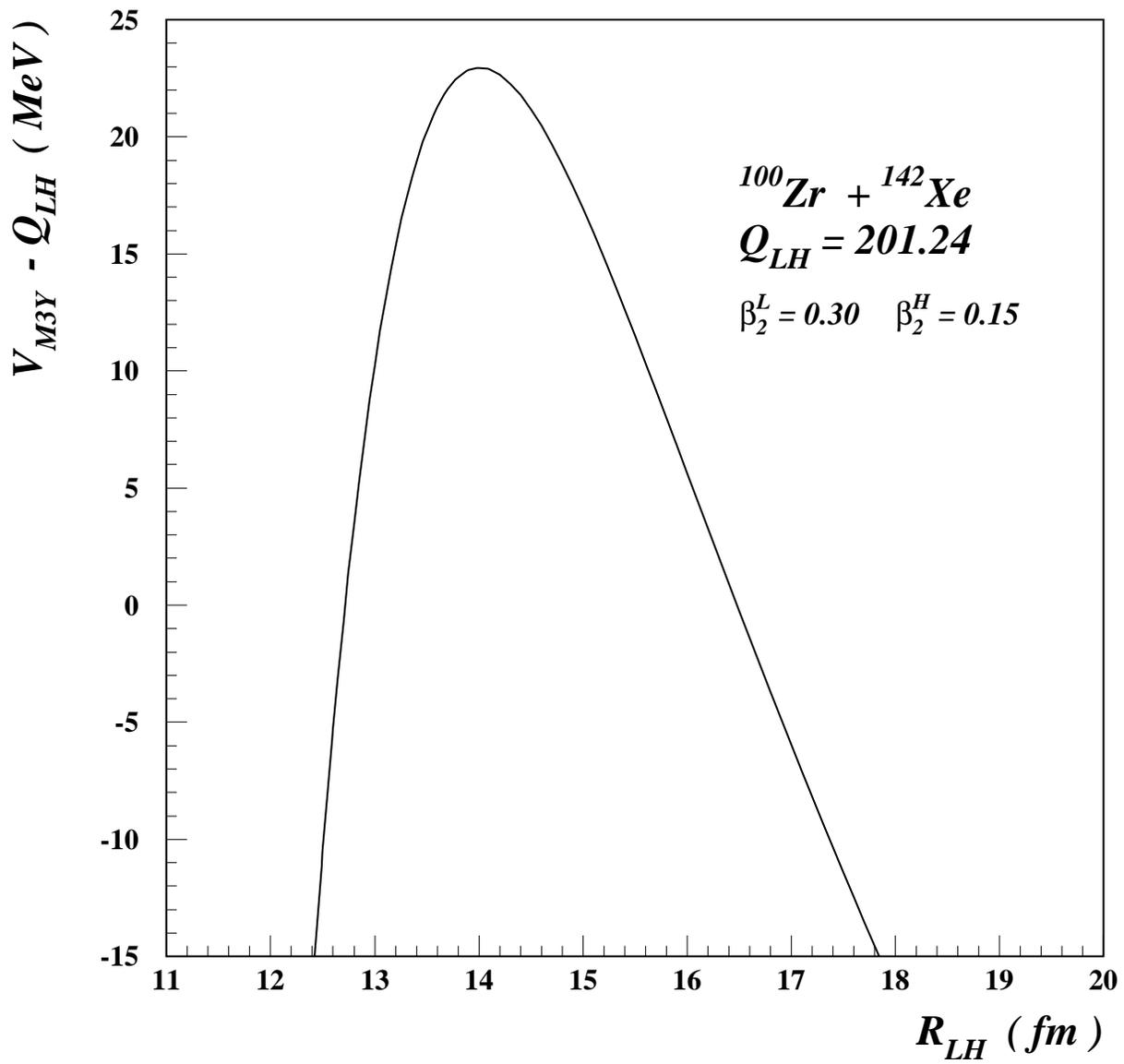
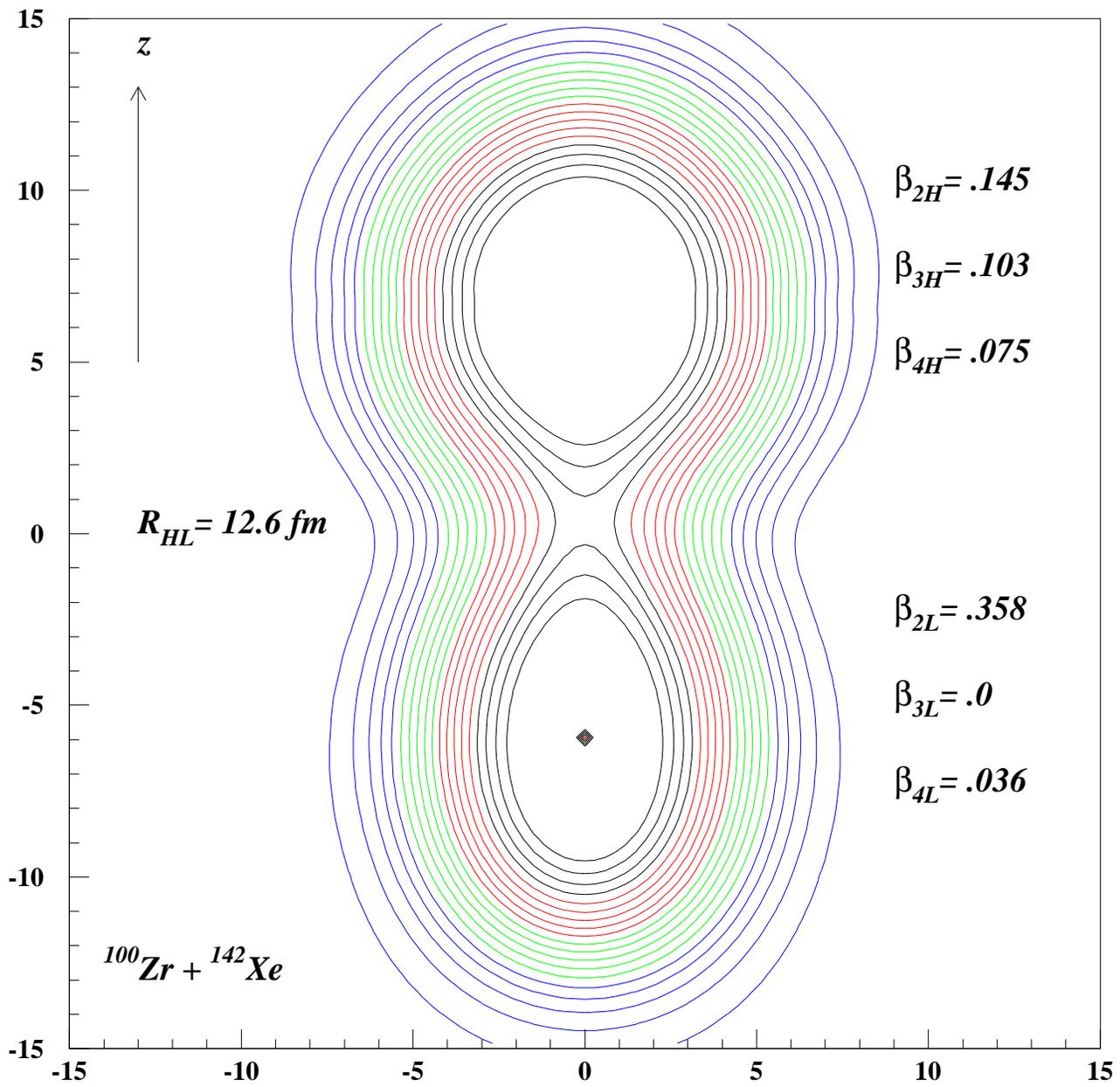
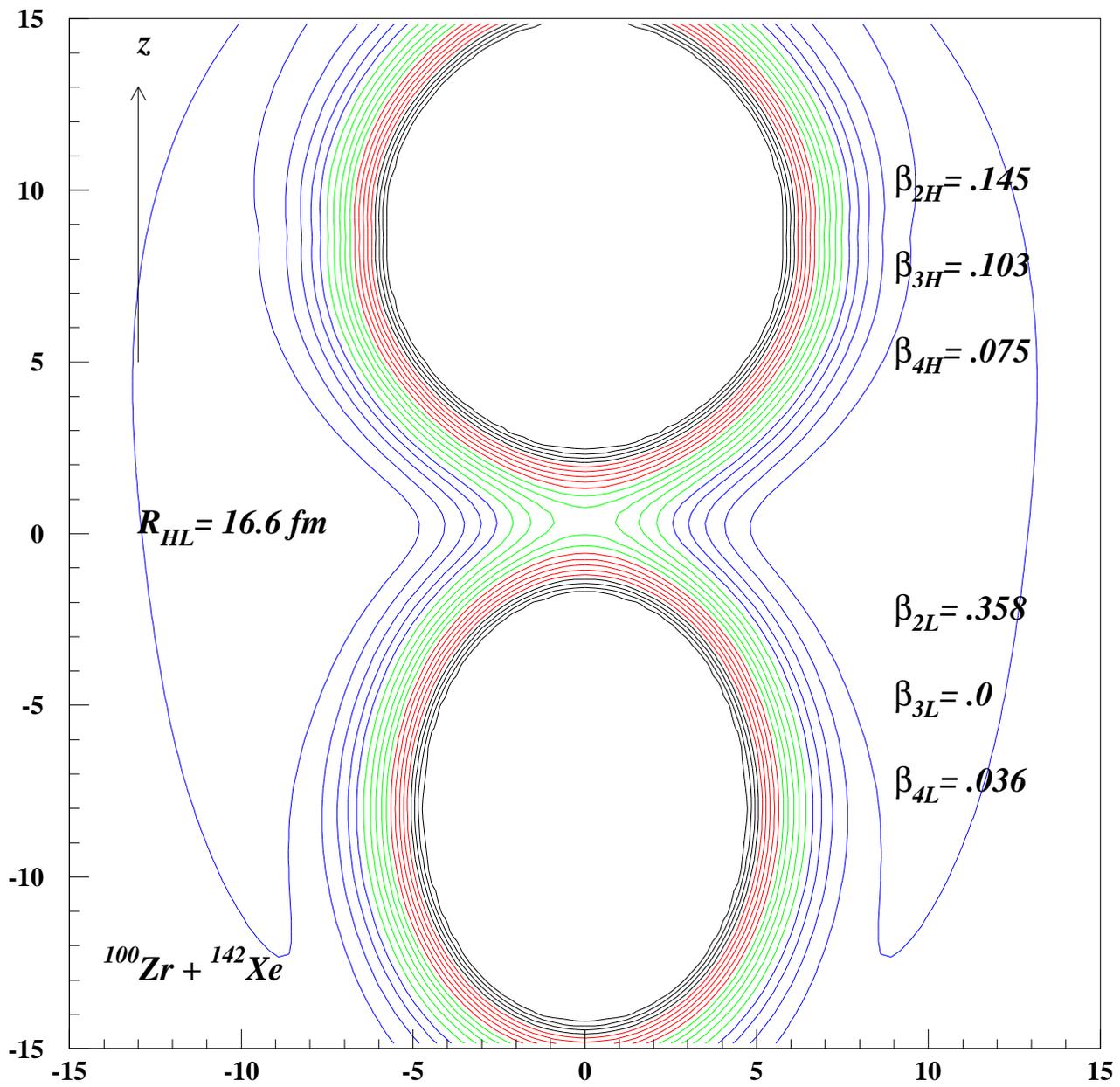
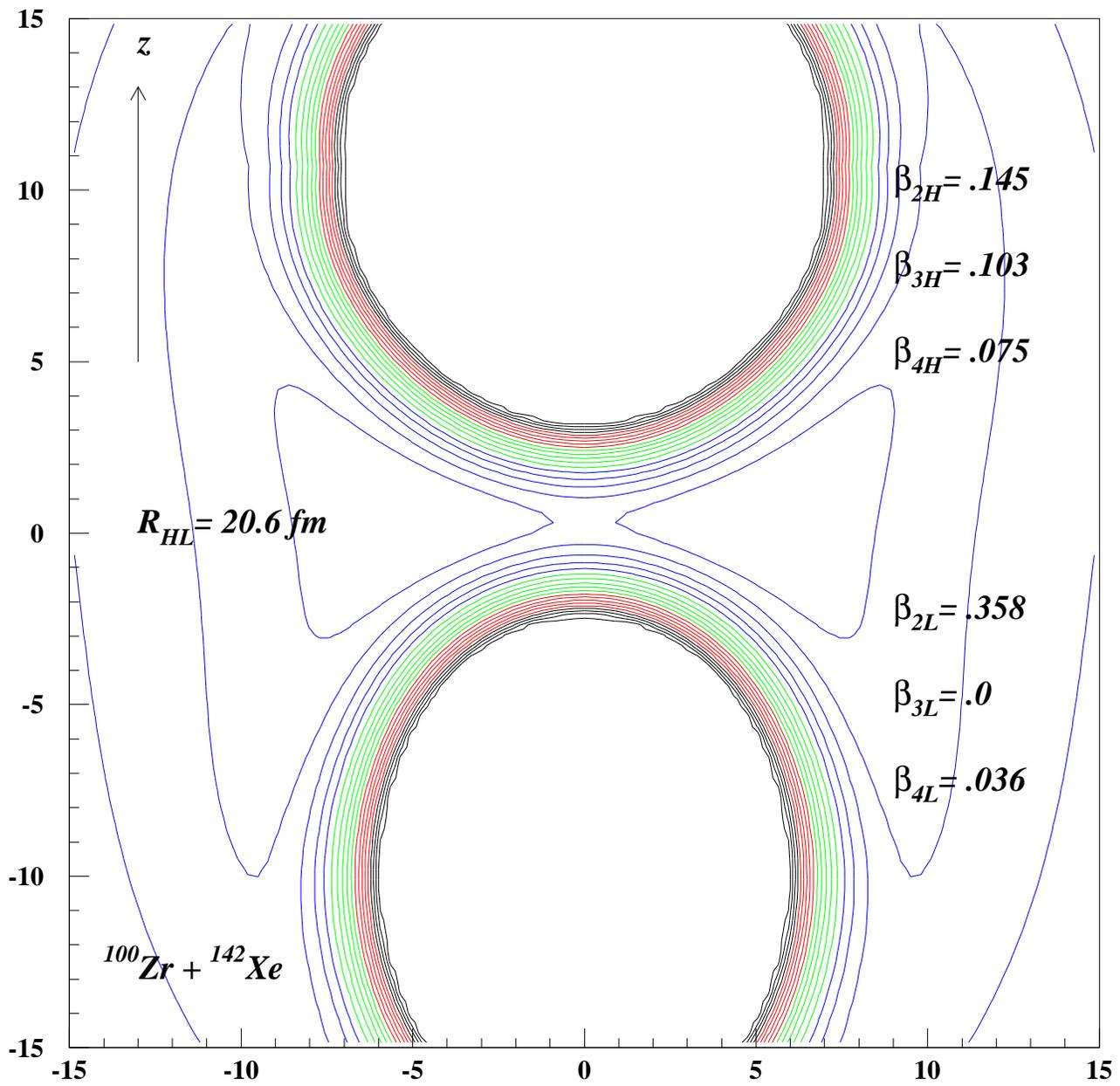
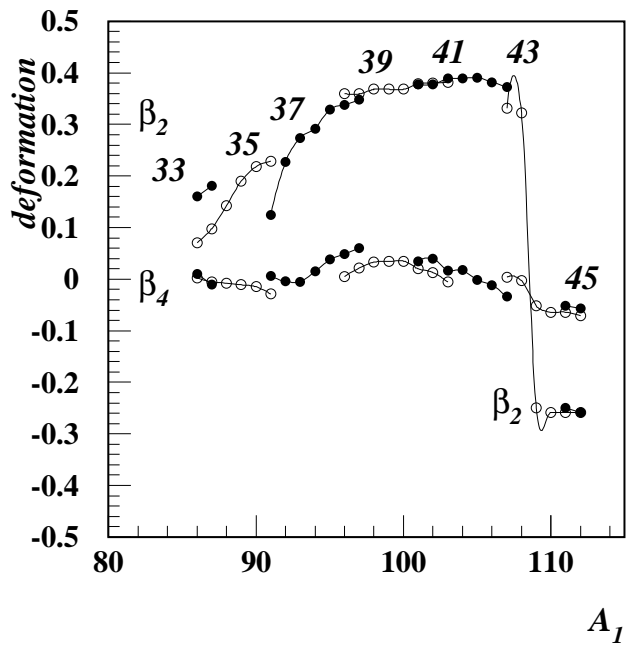
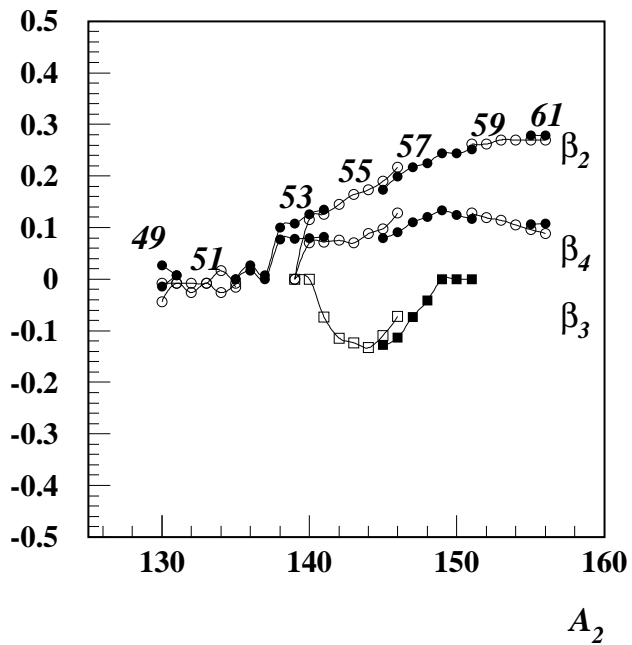
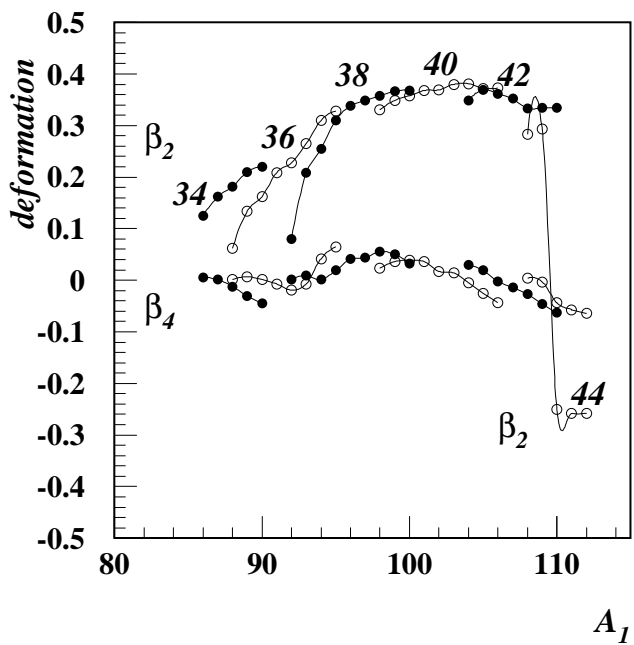
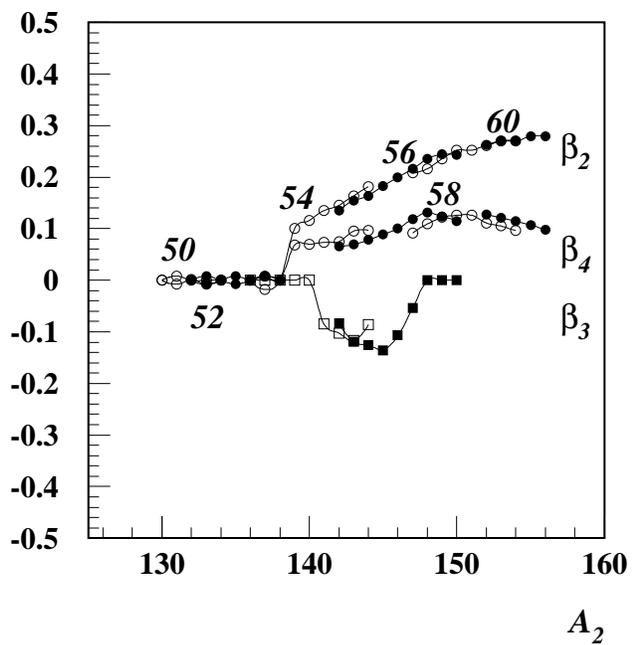
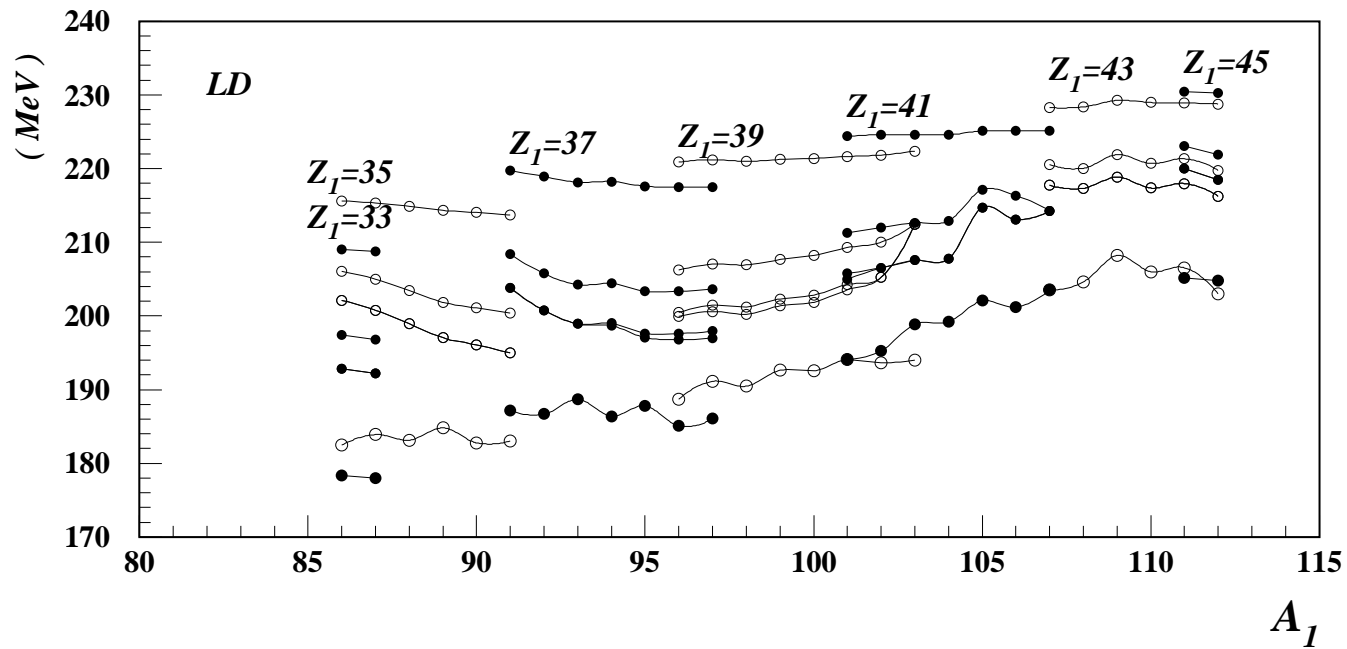
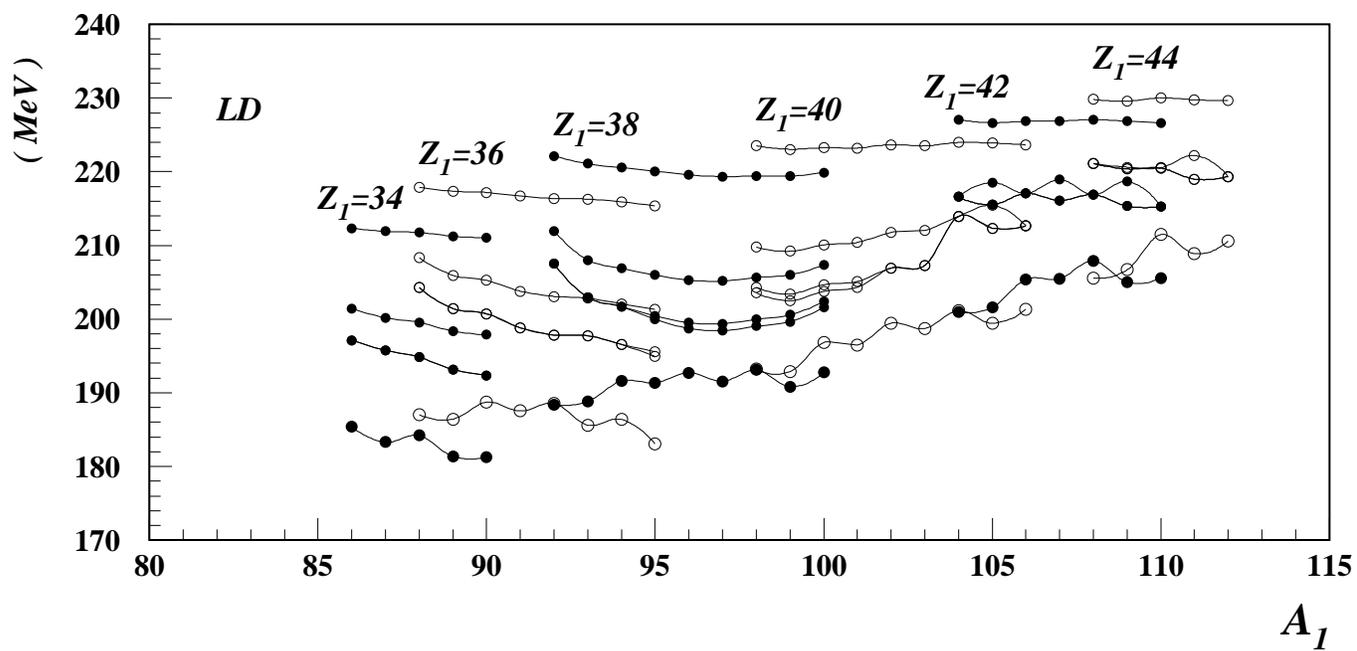
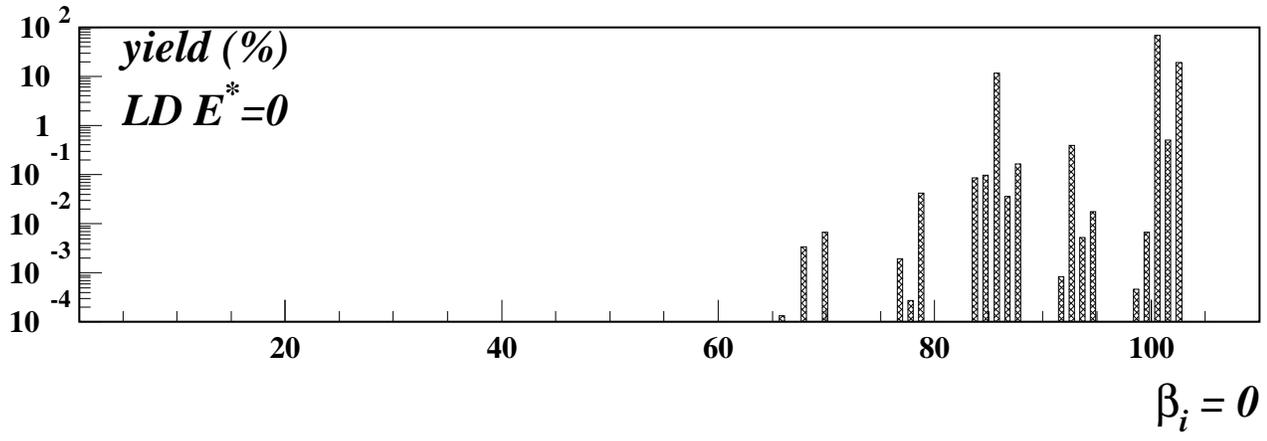
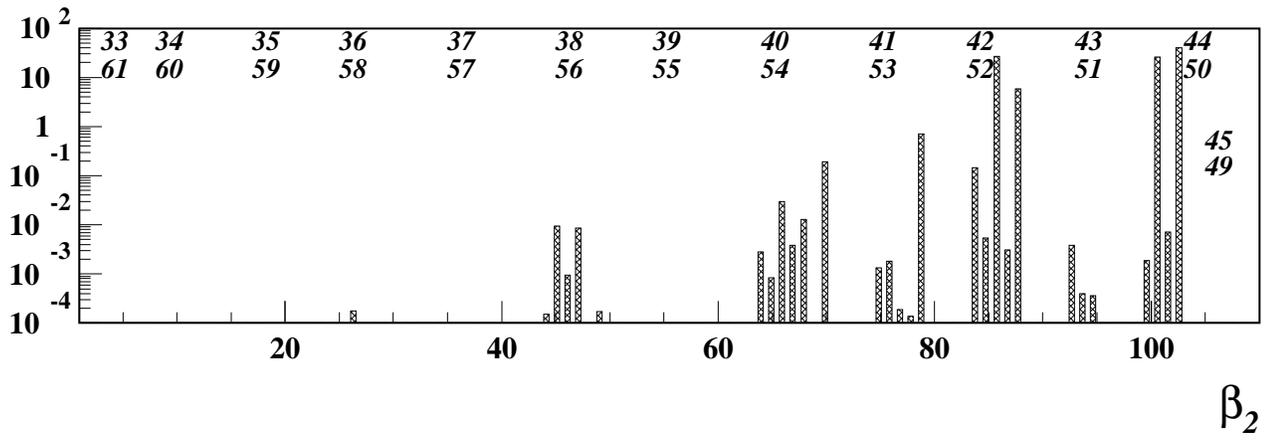
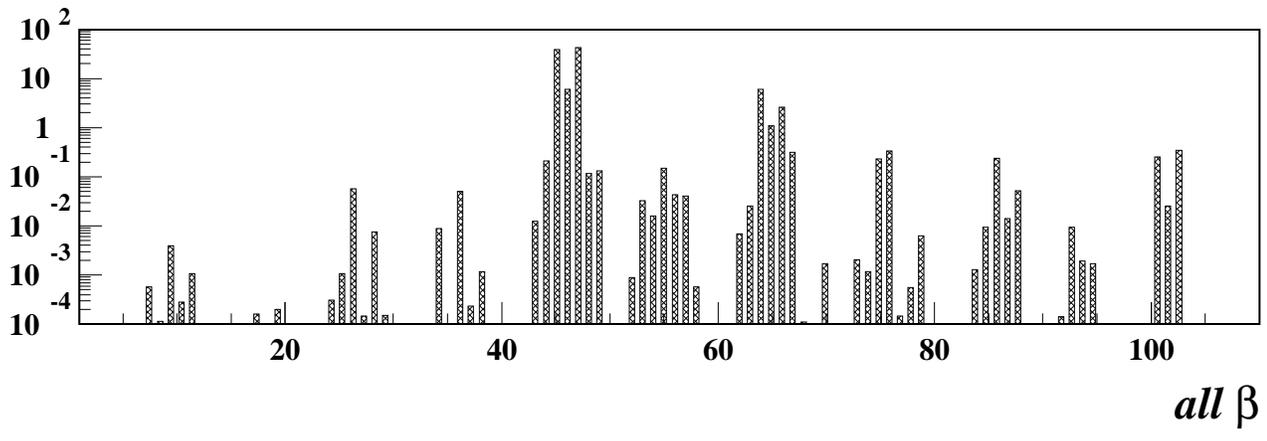
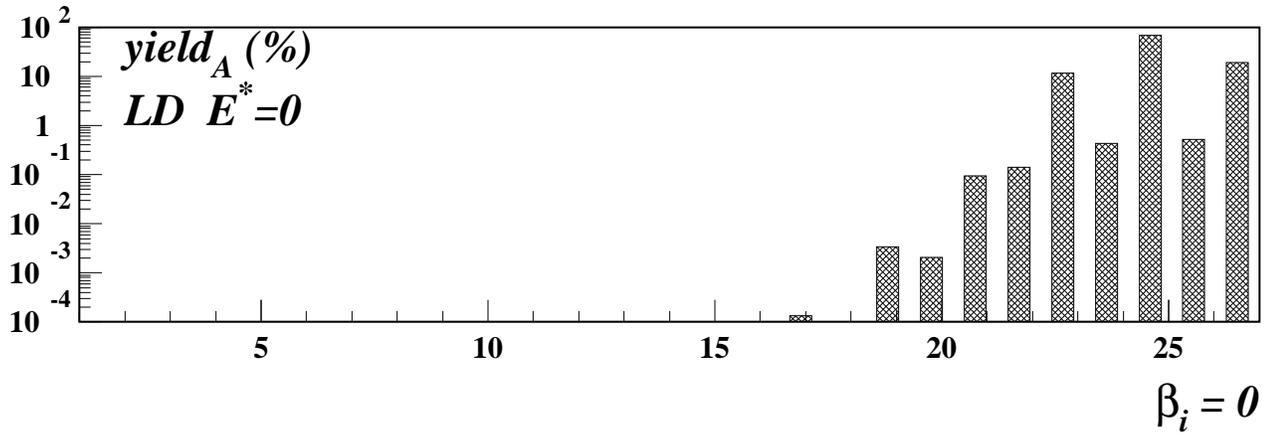
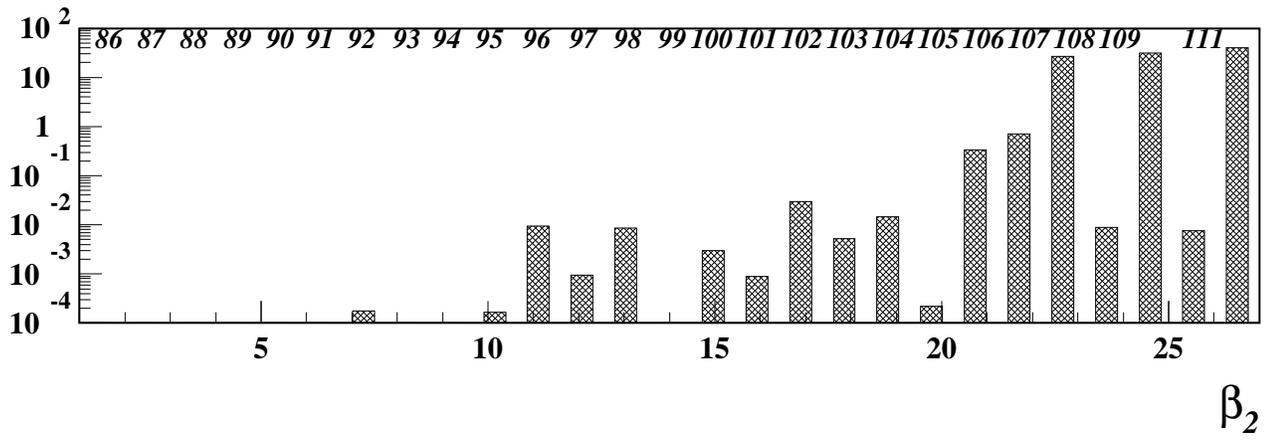
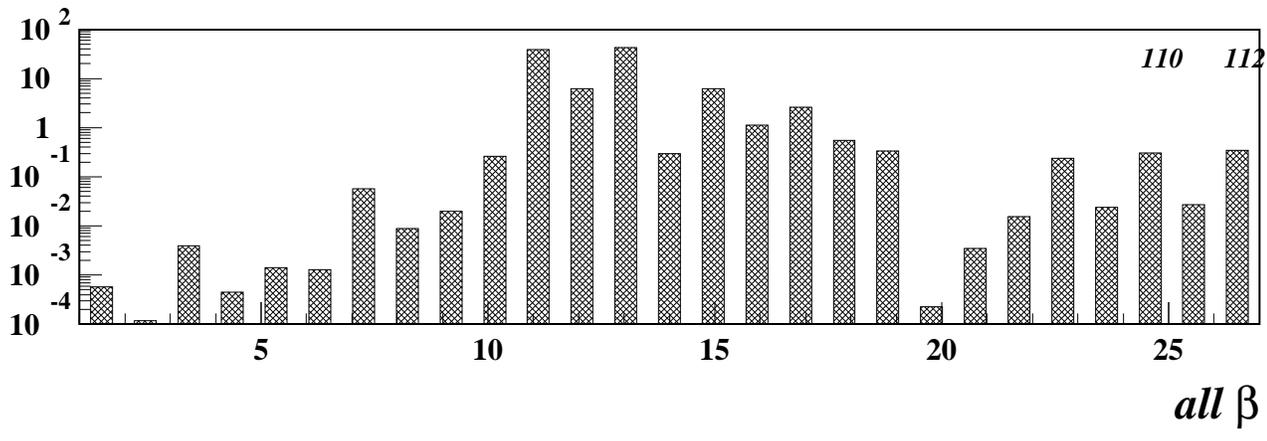
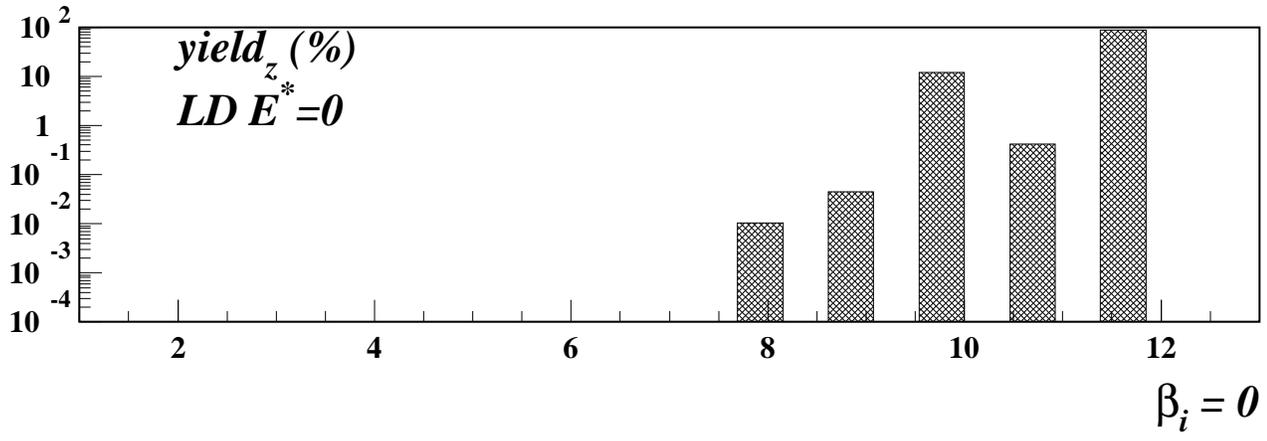
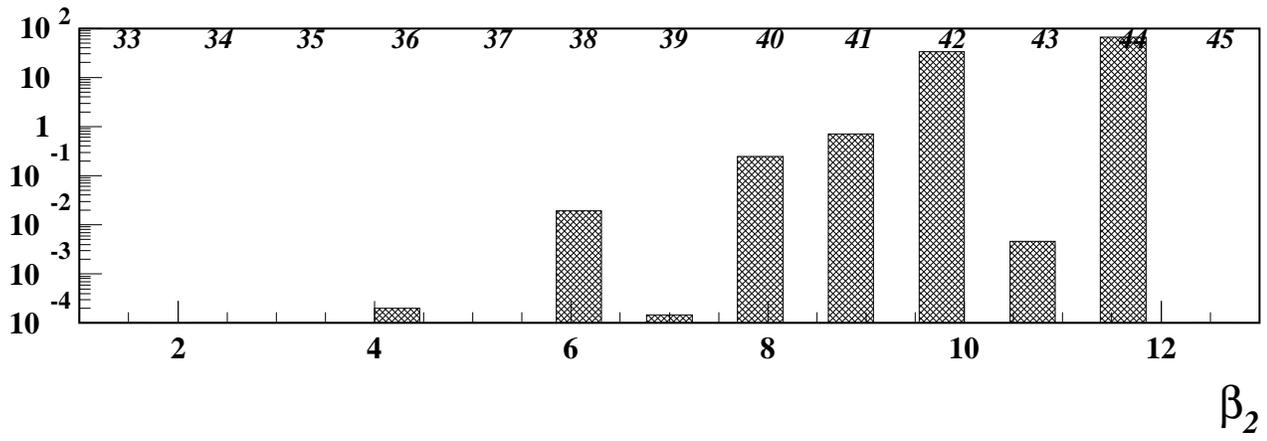
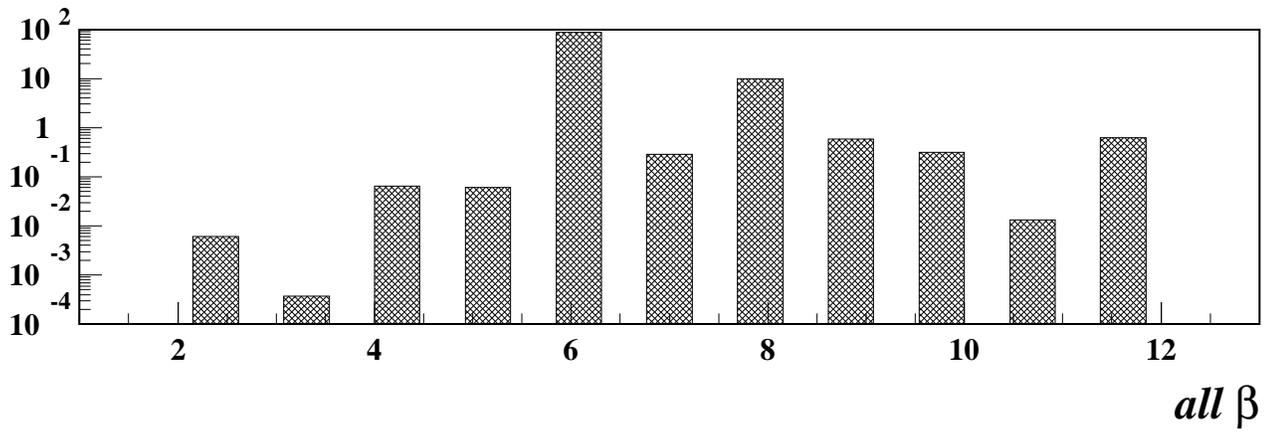
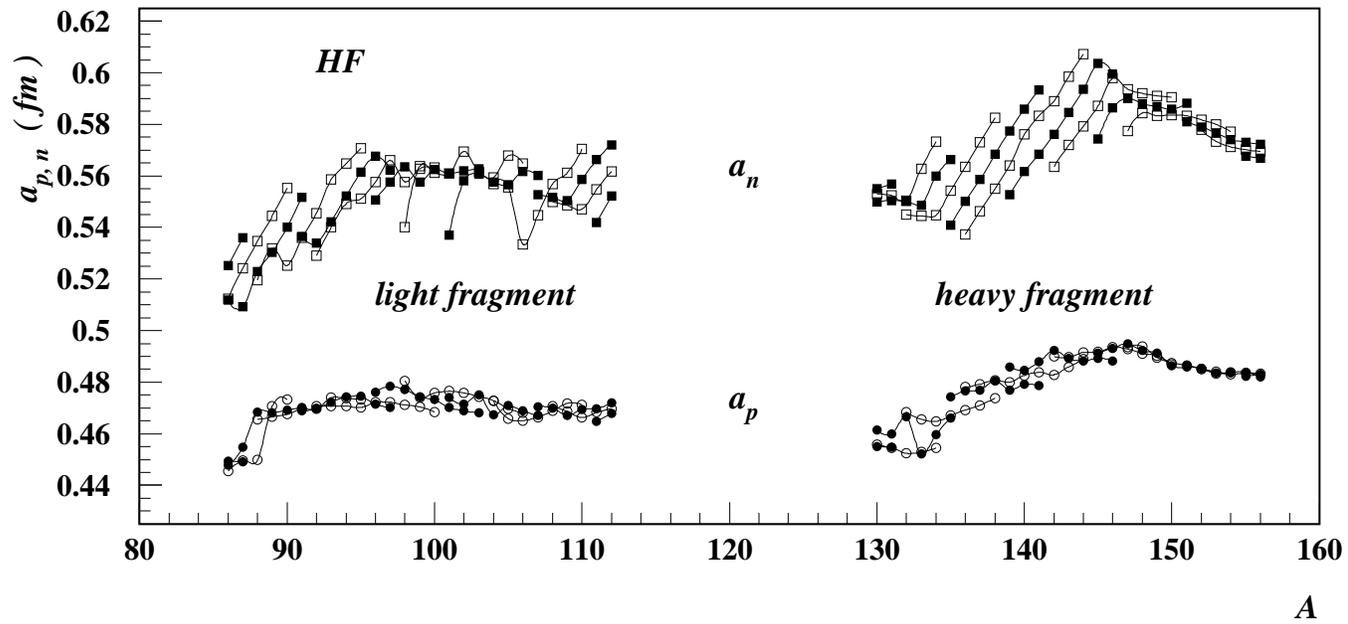
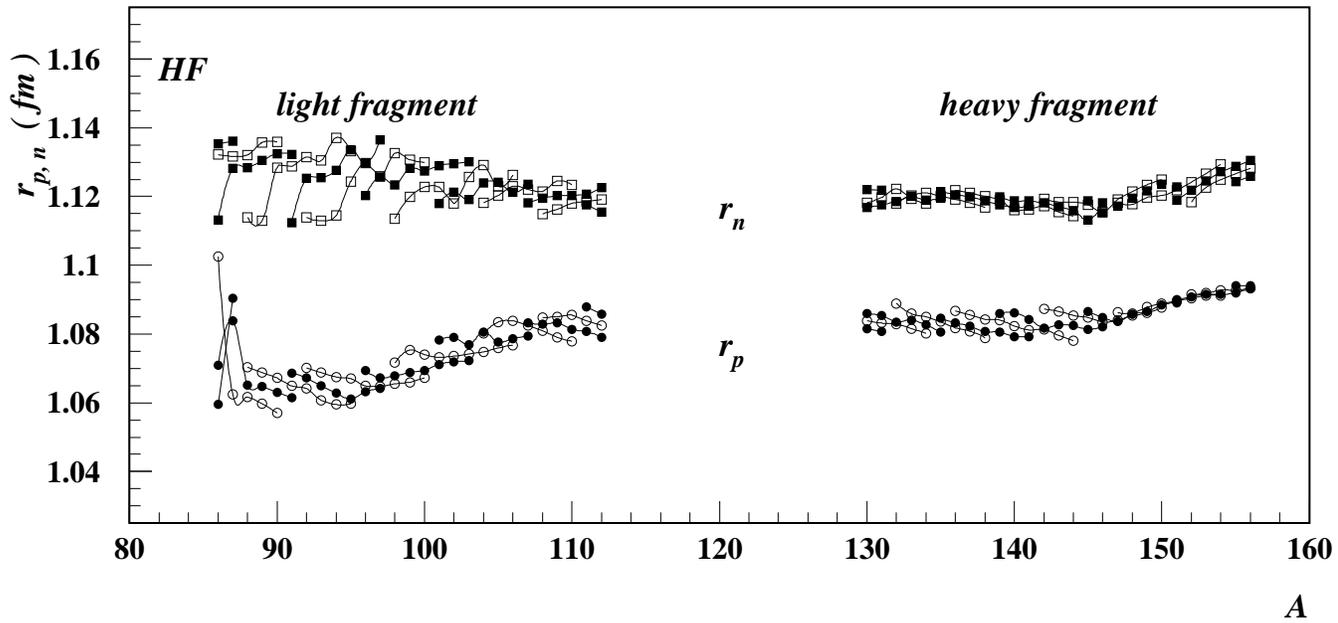
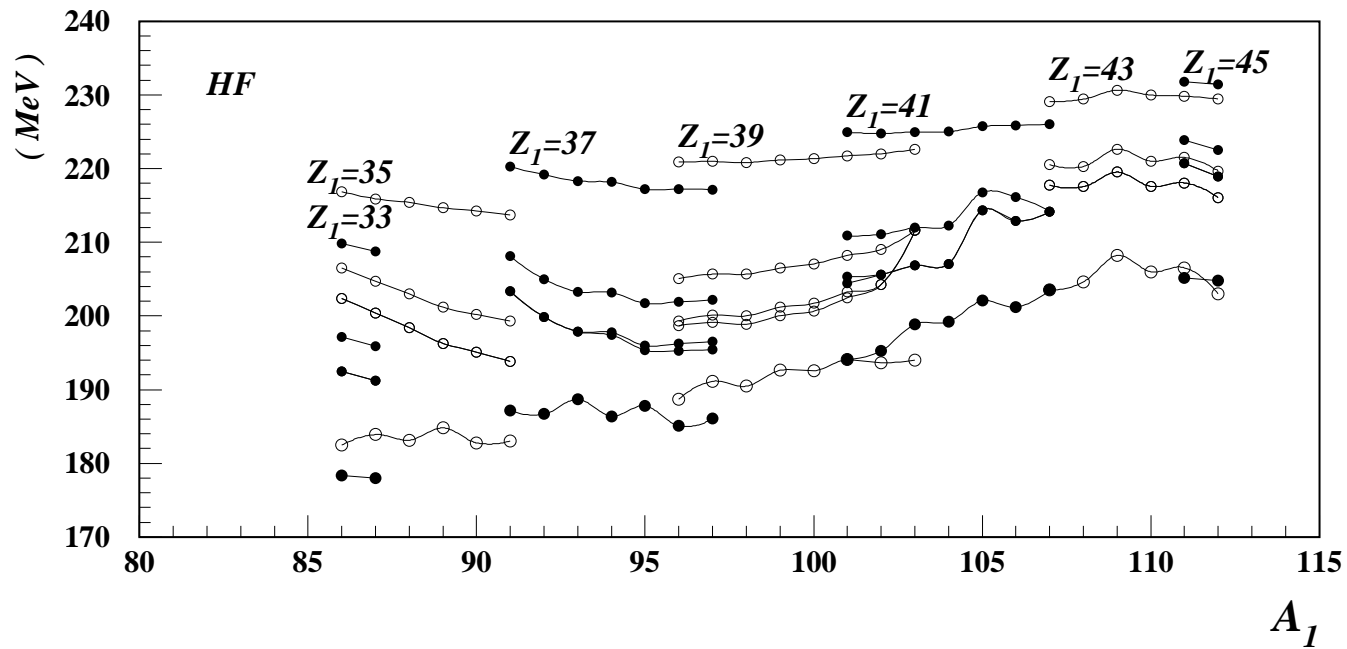
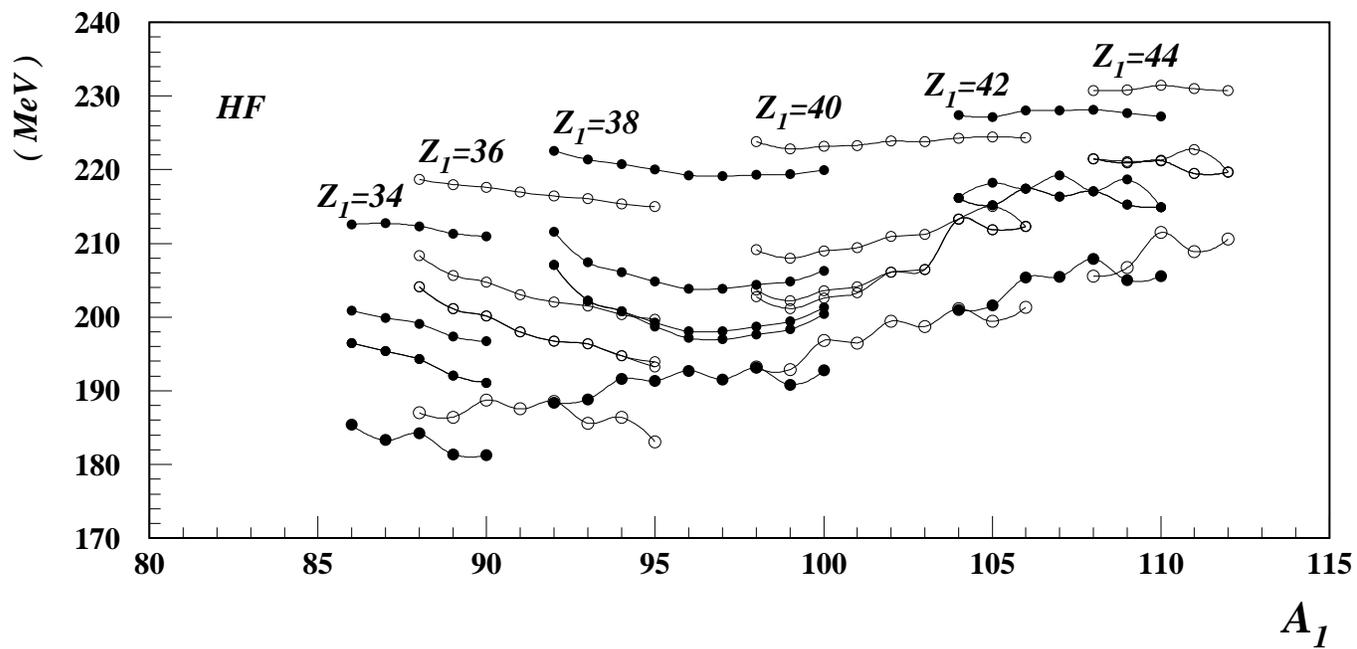
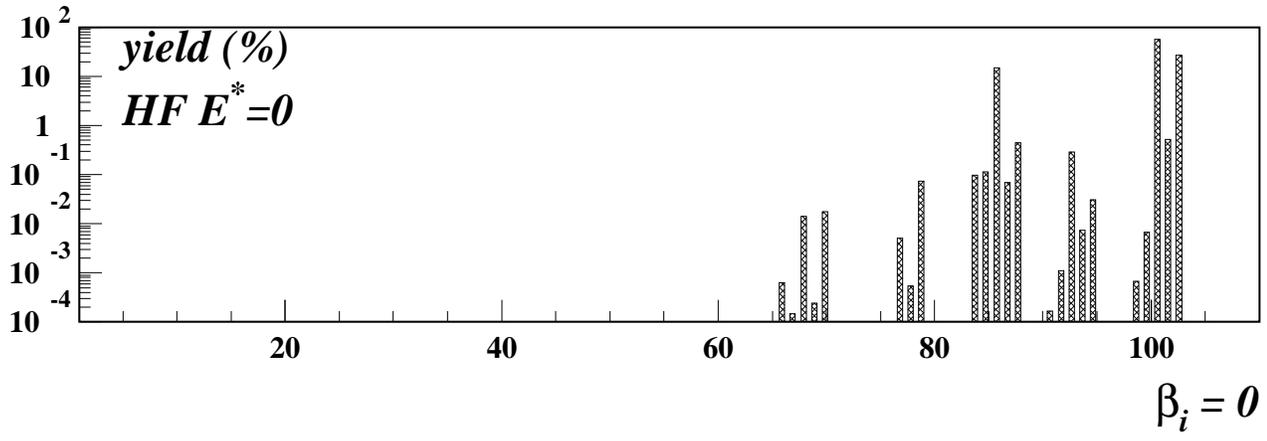
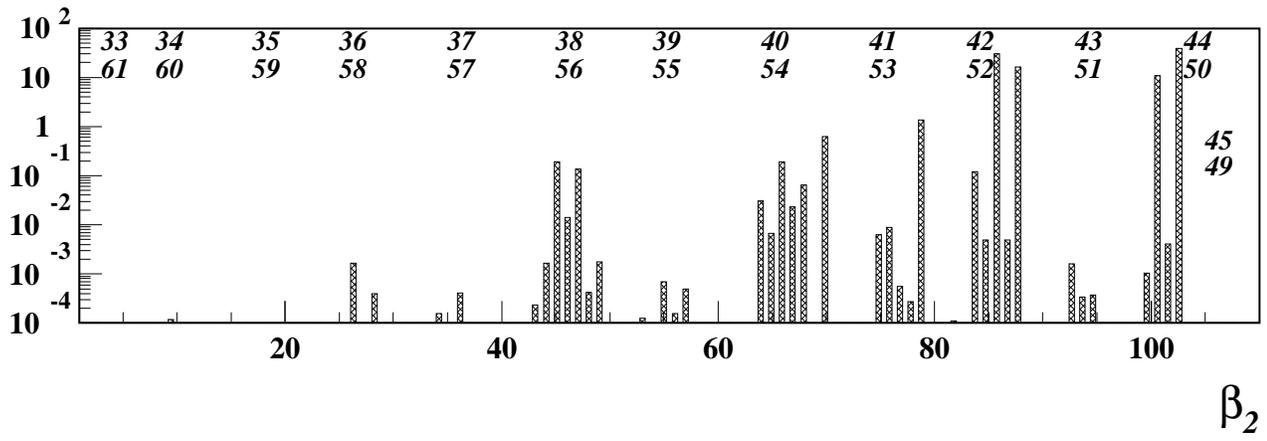
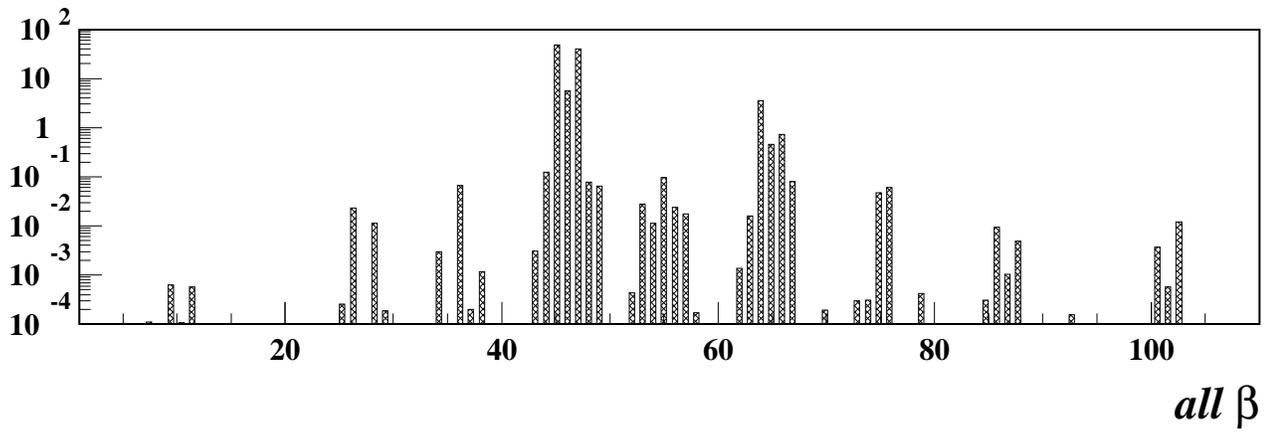
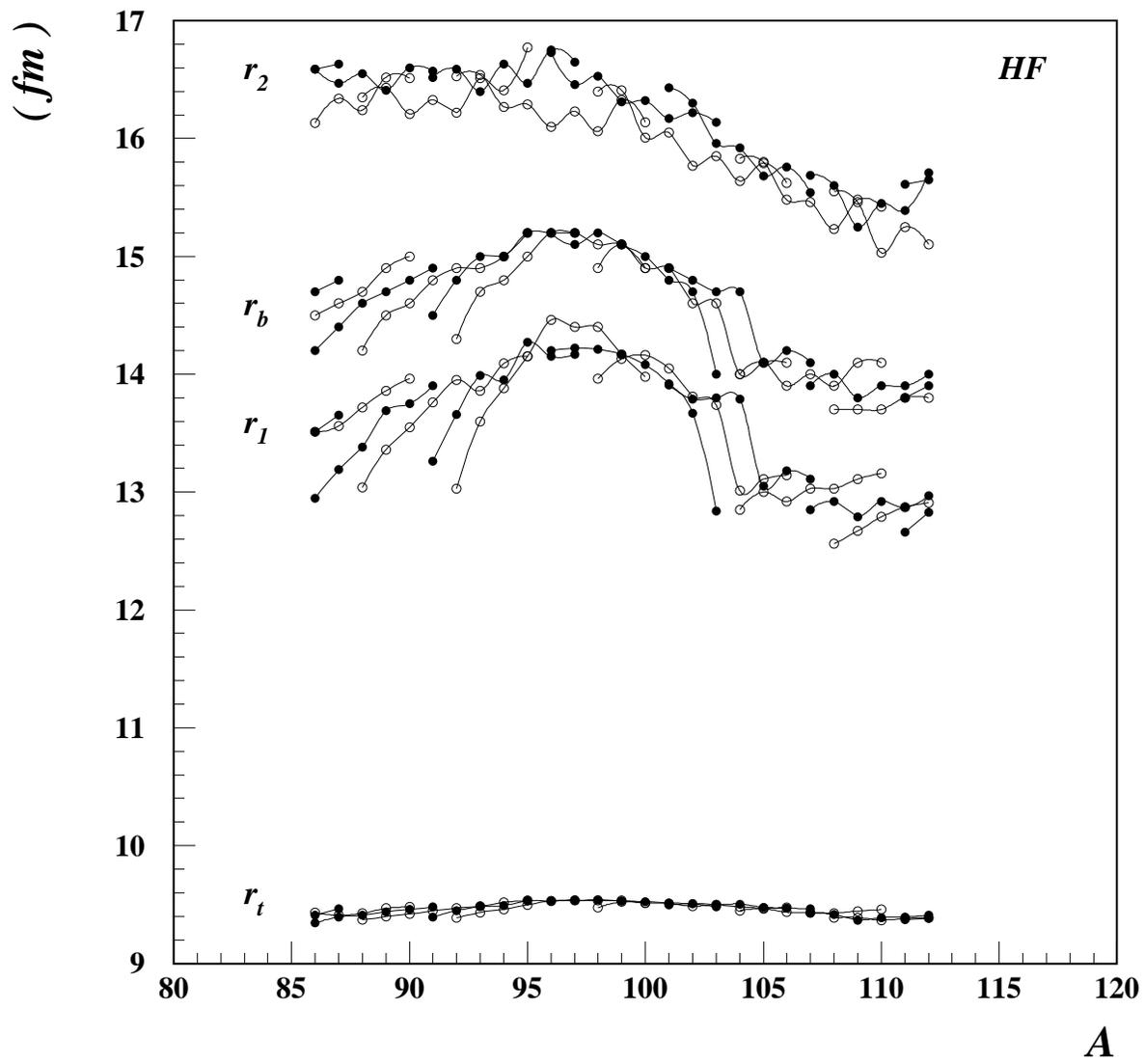
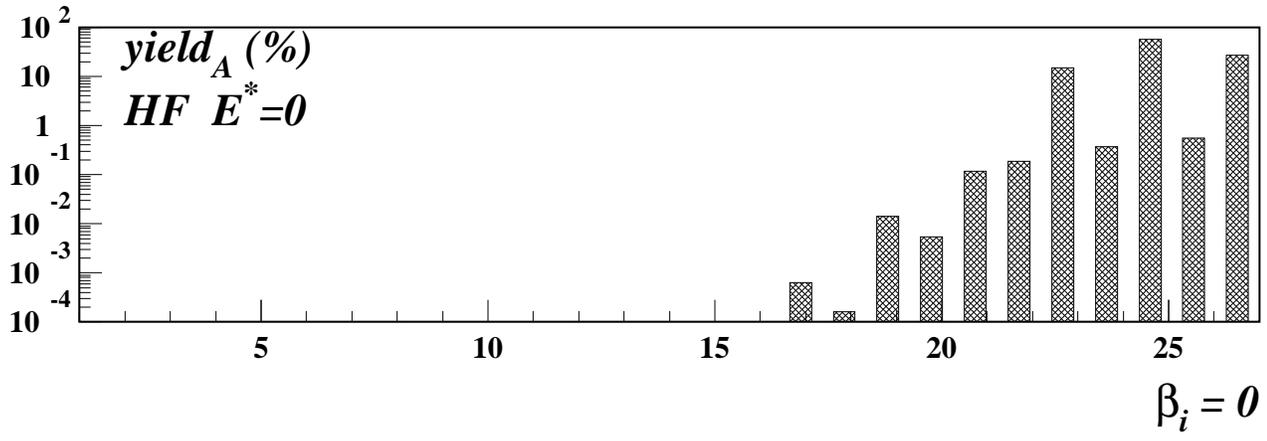
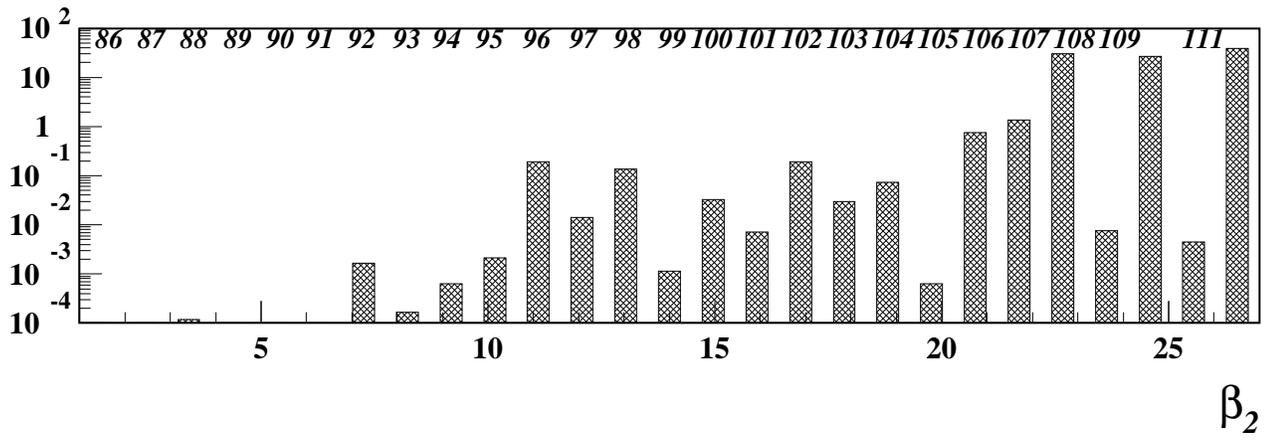
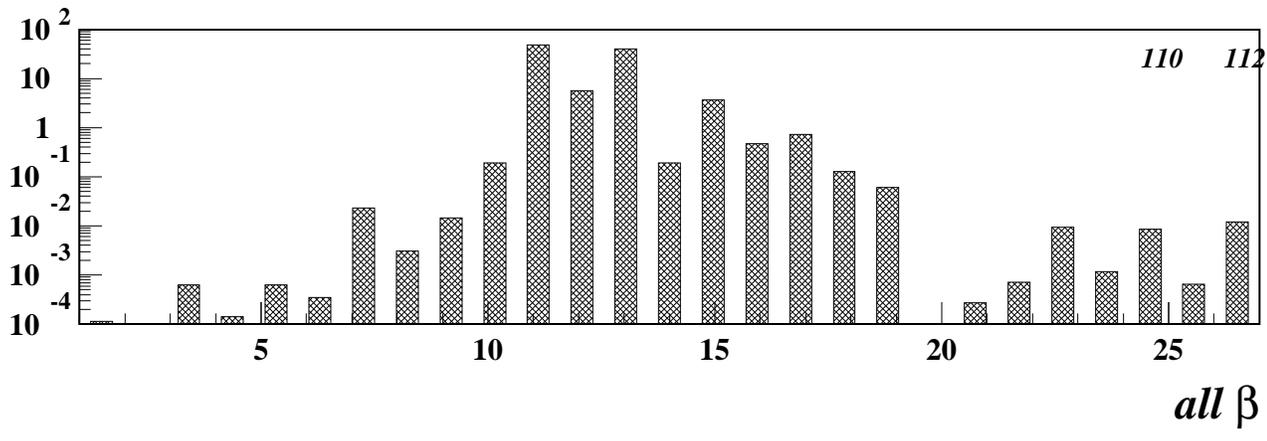
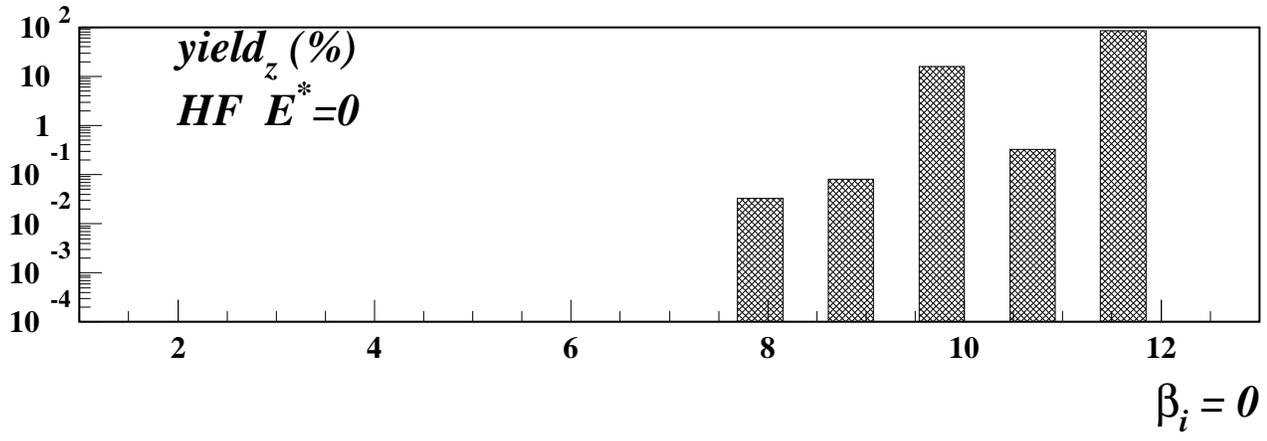
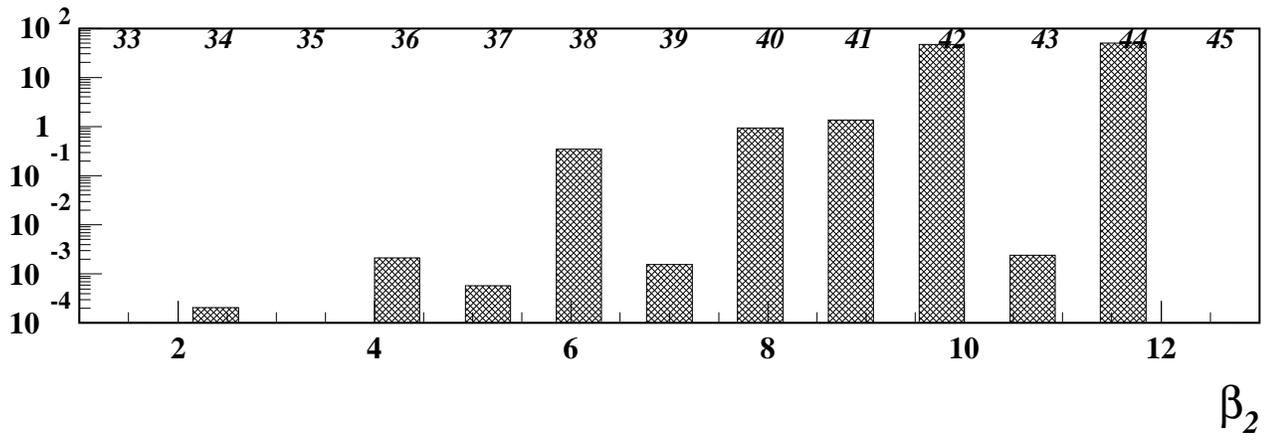
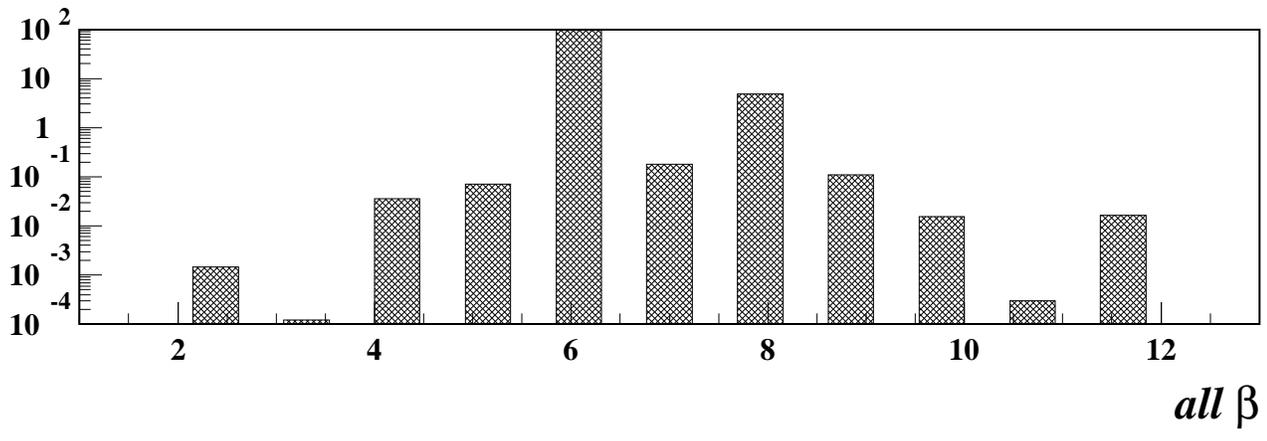
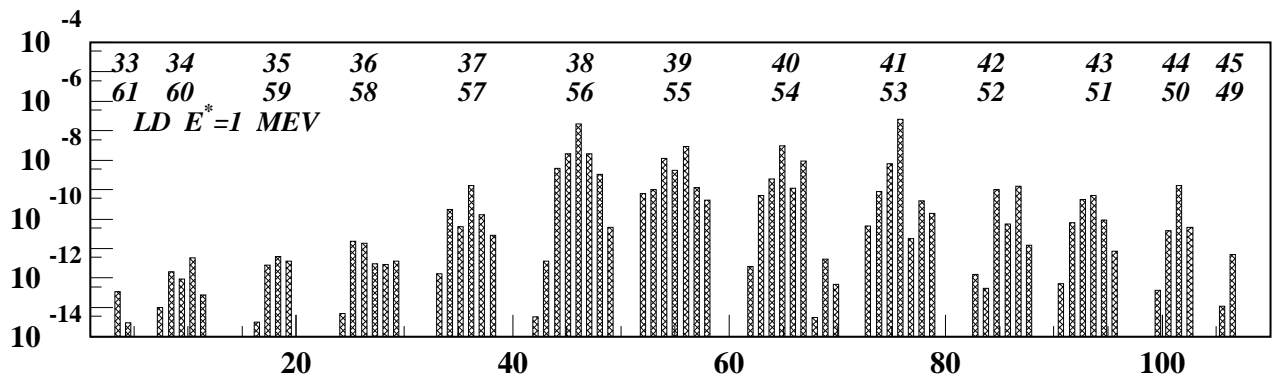
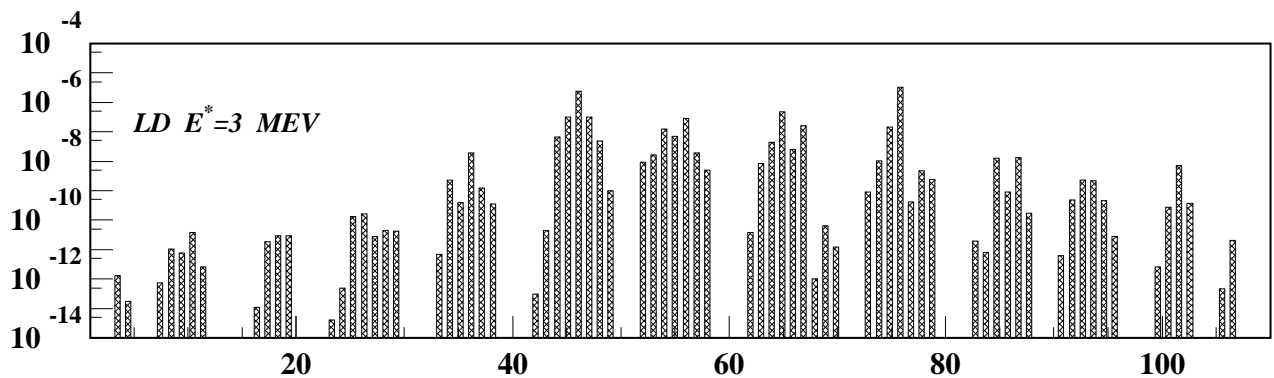
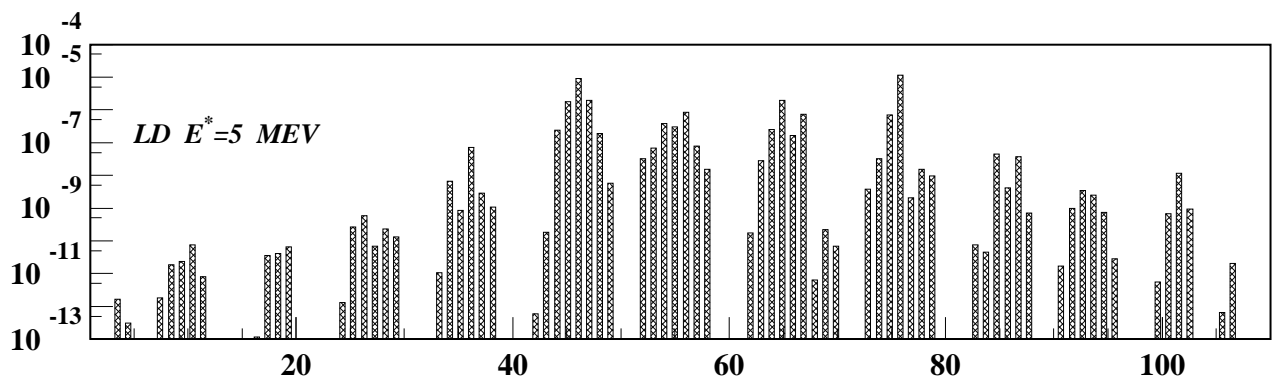
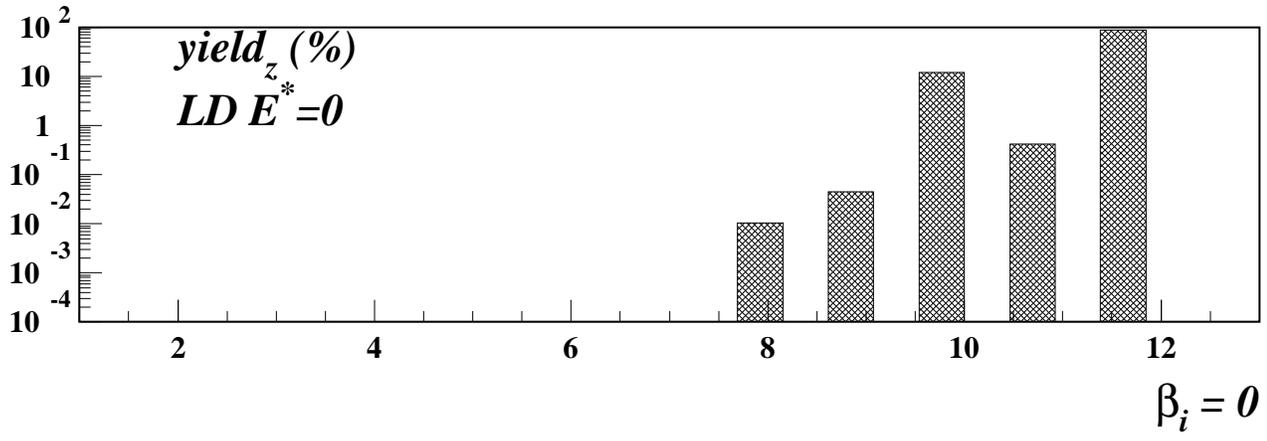
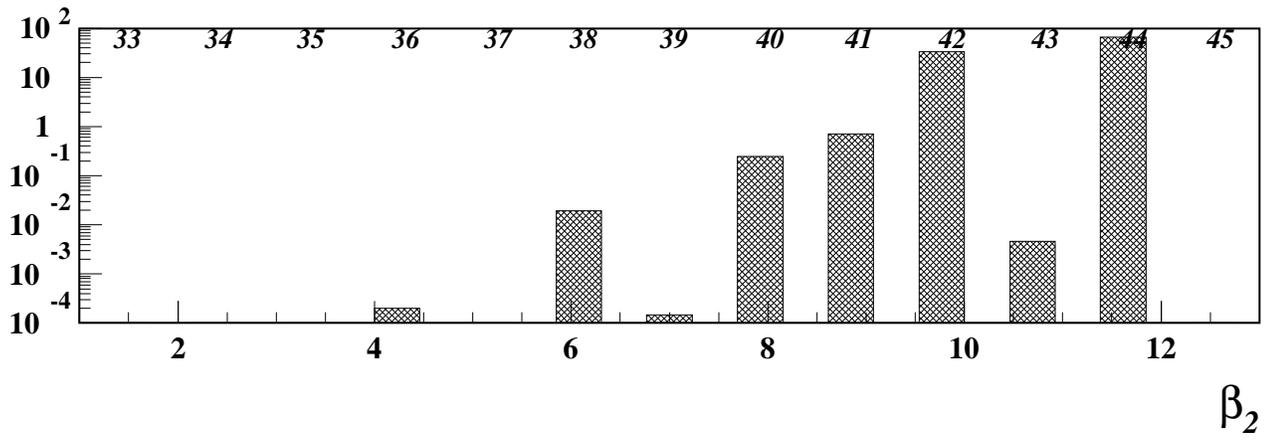
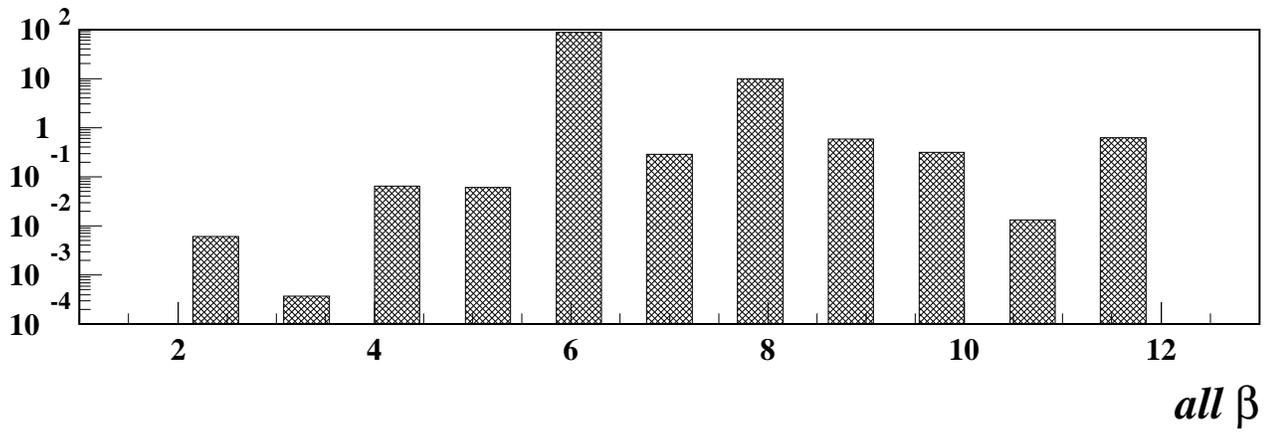
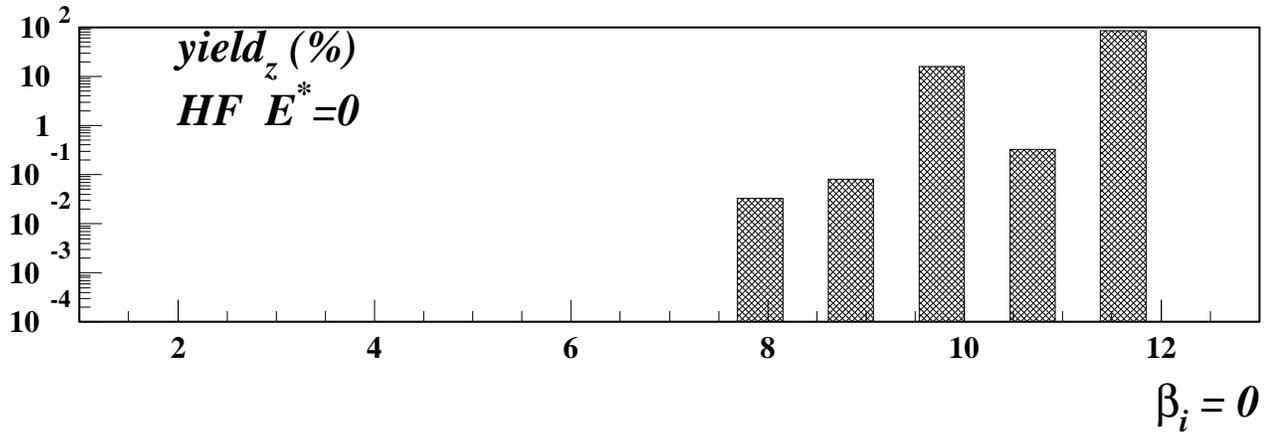
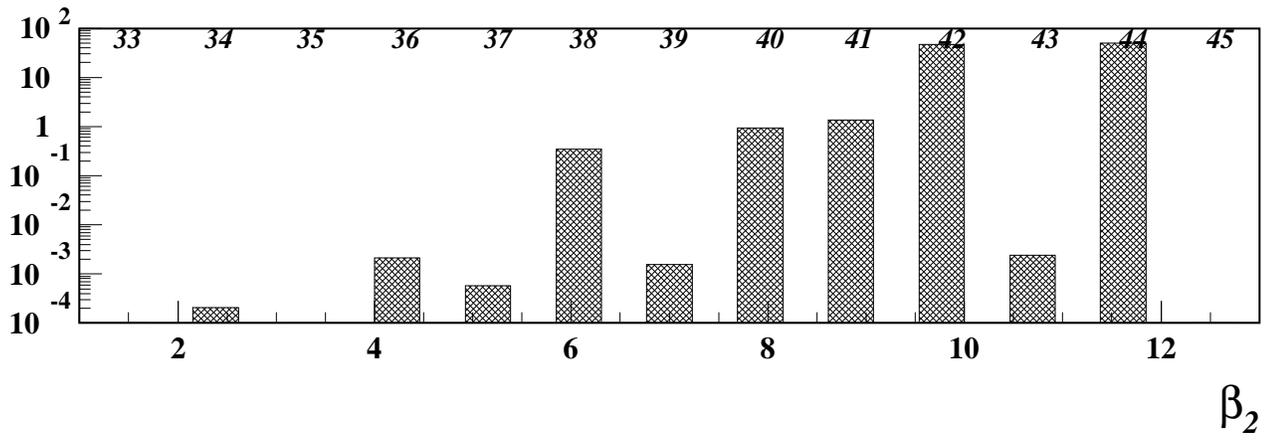
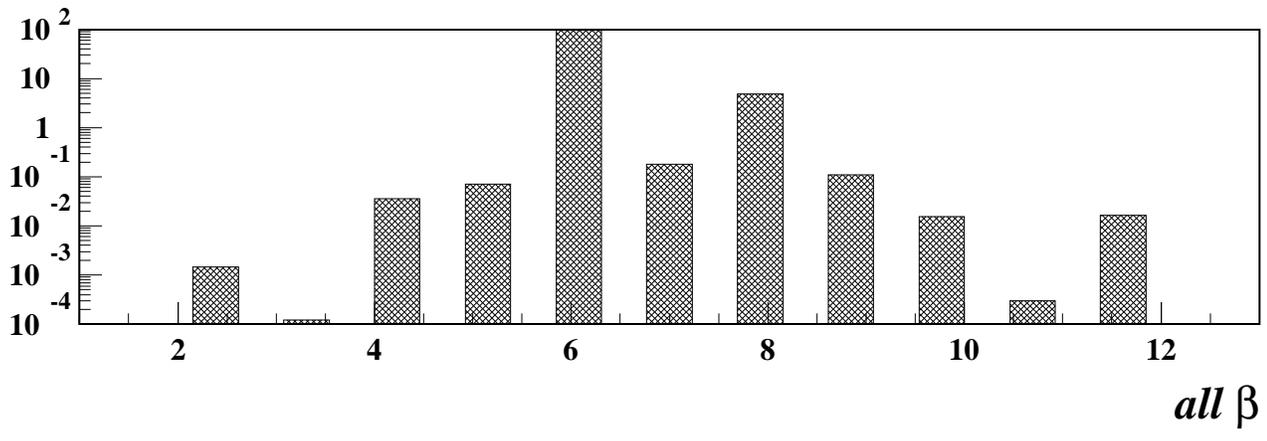
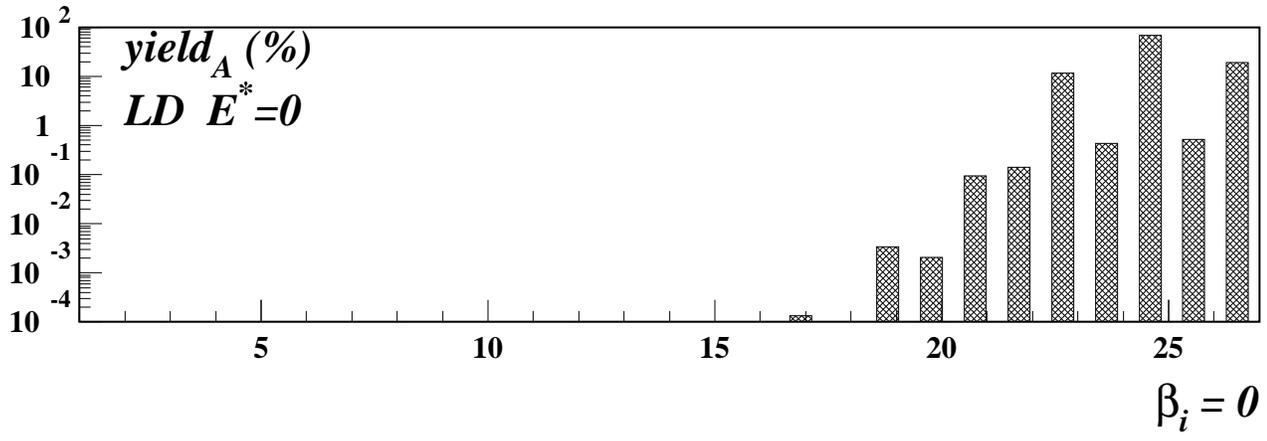
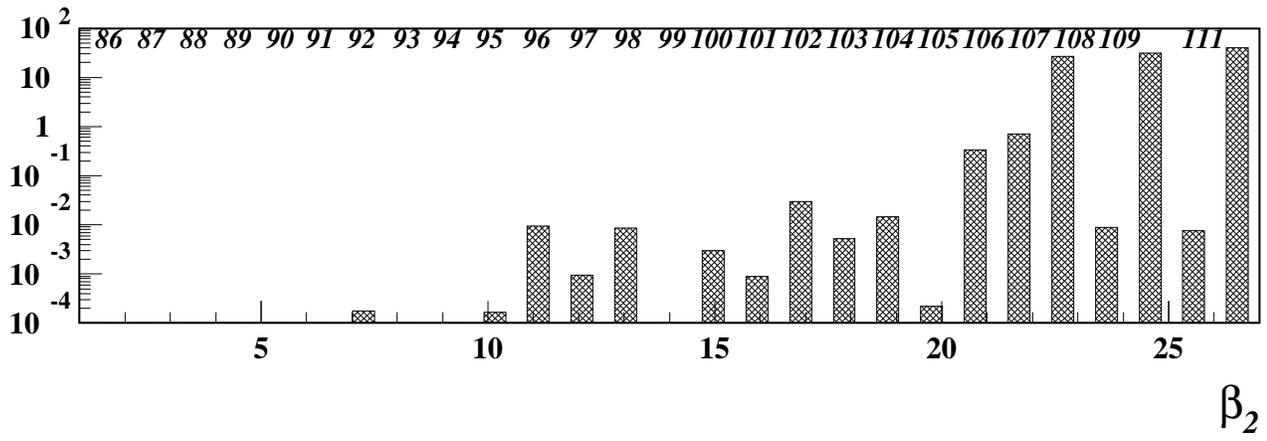
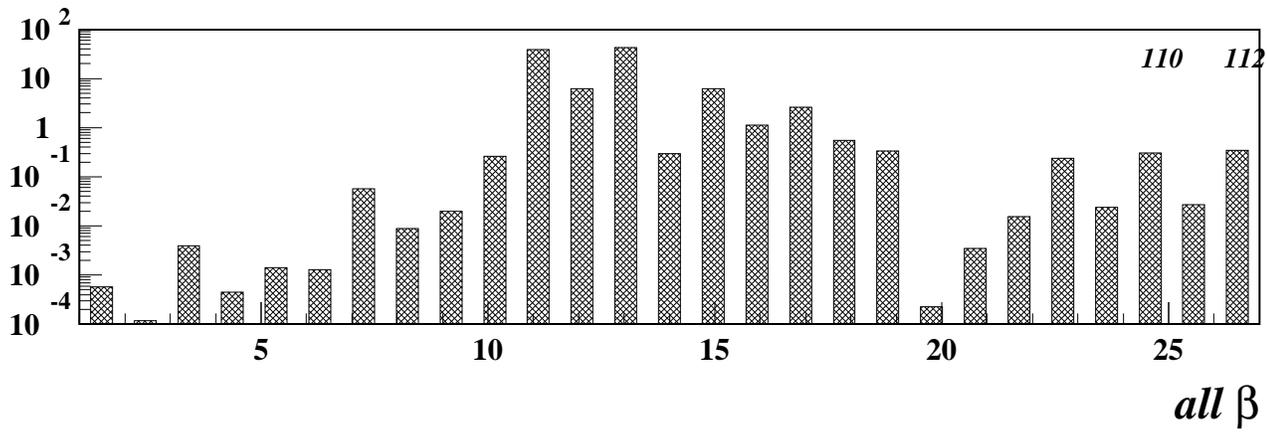
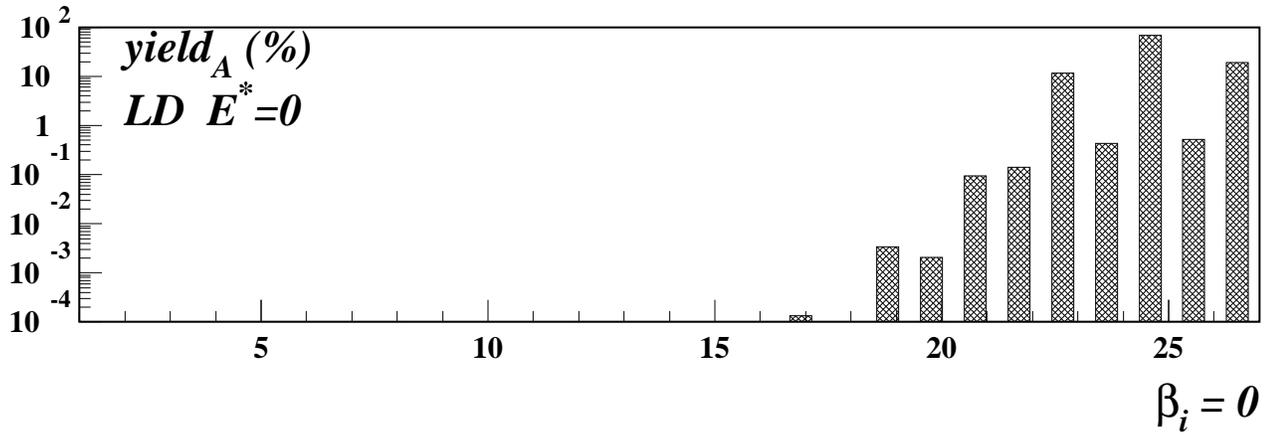
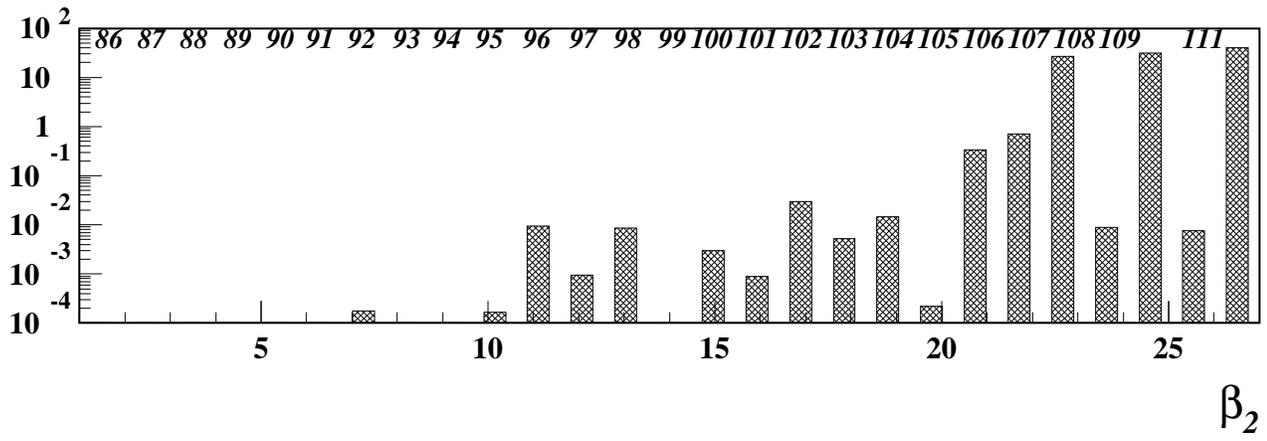
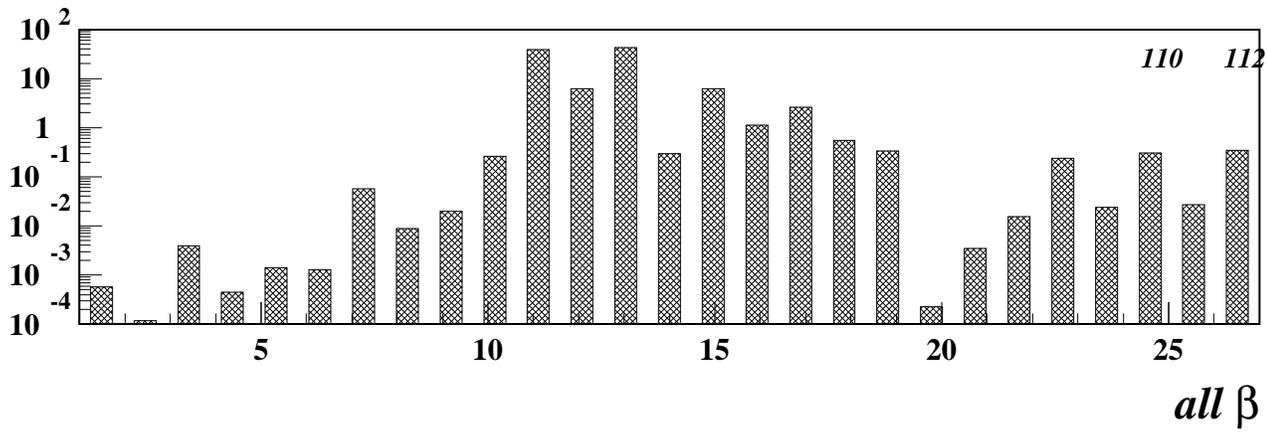
\end{document}